\documentclass[a4paper, table]{article}

\usepackage[utf8]{inputenc} 
\usepackage{amsmath, amsthm, amssymb, amsfonts} 
\usepackage{lipsum} 

\usepackage{graphicx} 
\usepackage{float} 
\usepackage{caption} 
\usepackage{subcaption} 

\usepackage{tabularx} 
\usepackage{tabu} 
\usepackage{booktabs} 

\usepackage[nottoc,numbib]{tocbibind} 
\usepackage[margin = 2.5cm]{geometry} 
\usepackage{microtype} 
\usepackage{titlesec} 
\usepackage{titletoc} 
\usepackage{appendix} 
\usepackage{fancyhdr} 
\usepackage[shortlabels]{enumitem} 

\usepackage{hyperref} 
\usepackage[noabbrev, capitalise]{cleveref} 

\usepackage{xcolor} 

\usepackage{tikz} 
\usetikzlibrary{positioning} 
\usetikzlibrary{calc} 

\makeatletter
\newcommand{\gettikzxy}[3]{%
  \tikz@scan@one@point\pgfutil@firstofone#1\relax
  \edef#2{\the\pgf@x}%
  \edef#3{\the\pgf@y}%
}
\makeatother

\newcommand\reporttitle{Mass dependence of overshooting \\ \bigskip beneath convective envelopes}

\newcommand\reportsubtitle{ 
}

\newcommand\groupnumber{
\textbf{}
}

\usepackage{orcidlink}
\newcommand\reportauthors{
J. Pratt (a) \orcidlink{0000-0003-2707-3616}, I. Baraffe (b)  \orcidlink{0000-0001-8365-5982}, M.-G. Dethero (c)\orcidlink{0000-0003-1617-3697}, \\ 
 M. Stuck (a)\orcidlink{0000-0003-1905-4474} , D.G. Vlaykov (d) \orcidlink{0000-0002-1402-9125}, T. Goffrey (e)  \orcidlink{0000-0003-0784-1294}
\vspace*{20mm}}

\newcommand\grouptutor{\small
\begin{enumerate}[a)]
\item Astronomy and Astrophysics Analytics Group, Lawrence Livermore National Laboratory, 7000 East Ave, Livermore, CA 94550, USA 
\item Astrophysics, College of Engineering, Mathematics and Physical Sciences, University of Exeter, EX4 4QL Exeter, United Kingdom 
\item Department of Physics and Astronomy, Georgia State University, Atlanta GA 30303, USA 
\item Geophysical and Astrophysical Fluid Dynamics, Department of Mathematics and Statistics, Univerity of Exeter, EX4 4QE, Exeter, United Kingdom
\item Centre for Fusion, Space and Astrophysics, Department of Physics, University of Warwick, Coventry, CV4 7AL, United Kingdom
\end{enumerate}
}

\newcommand\placeanddate{
Livermore, California \today
}

\definecolor{Tue-red}{RGB}{199, 25, 24}
\definecolor{lightblue}{rgb}{.8, 1., 1.}
\definecolor{cadet}{rgb}{.3725, .619, .627}
\definecolor{cyan}{rgb}{0., .545, .545}
\definecolor{sea}{rgb}{.235, .702, .443}
\definecolor{aqua}{rgb}{.561, .737, .561}
\definecolor{turq}{rgb}{.686, .9333, .9333}
\definecolor{whiteblue}{rgb}{0.2, .8, .6}
\definecolor{bluey}{rgb}{0.2, .8, 1.}
\definecolor{ltblue4}{rgb}{.902, .957, 1}
\definecolor{seablue}{rgb}{.3725,  .619,  .627}
\definecolor{dodger}{rgb}{.0,.2758,.5151}

\hypersetup{
    colorlinks=true,
    linkcolor=dodger,
    urlcolor=dodger,
    citecolor = dodger
    }
\counterwithin{equation}{section} 
\counterwithin{figure}{section} 
\counterwithin{table}{section} 

\titleformat{\section}{\sffamily\color{dodger}\Large\bfseries}{\thesection\enskip\color{gray}\textbar\enskip}{0cm}{} 

\titleformat{\subsection}{\sffamily\color{dodger}\large\bfseries}{\thesubsection\enskip\color{gray}\textbar\enskip}{0cm}{} 

\titleformat{\subsubsection}{\sffamily\color{dodger}\bfseries}{\thesubsubsection\enskip\color{gray}\textbar\enskip}{0cm}{} 

\DeclareCaptionFont{dodger}{\color{dodger}}
\usepackage{mlmodern}

\usepackage{graphicx}
\usepackage{xspace}
\usepackage{amsmath}
\usepackage{amsfonts}
\usepackage{amssymb}
\usepackage{placeins}
\usepackage{gensymb}
\usepackage{color}
\usepackage{float}
\usepackage{hyperref}
\usepackage[english]{babel}
\usepackage{natbib}
\usepackage[document]{ragged2e}
\usepackage[skip=10pt plus 10pt, indent=10pt]{parskip}

\renewcommand{\vec}[1]{\mbox{\boldmath$#1$}}
\newcommand{\music}{\texttt{MUSIC}\xspace}

\begin{document}

\begin{titlepage}

\centering

\begin{tikzpicture}

\node[opacity=0.2,inner sep=0pt,remember picture,overlay] at (4.5,-0.5){\includegraphics[width= 0.8 \textwidth]{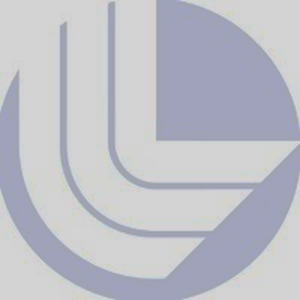}};

\node[inner sep=0pt] (logo) at (0,0)
    {\includegraphics[width=.25\textwidth]{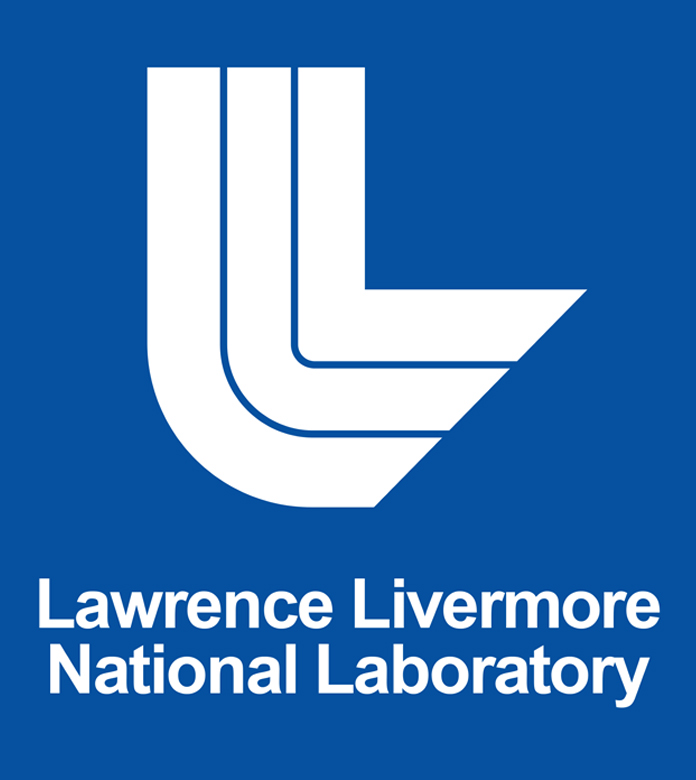}};
    
\node[text width = 0.7\textwidth, right = of logo](title){\sffamily\huge\reporttitle};

\node[text width = 0.5\textwidth, yshift = 0.75cm, below = of title](subtitle){\sffamily\Large \reportsubtitle};

\gettikzxy{(subtitle.south)}{\sffamily\subtitlex}{\subtitley}
\gettikzxy{(title.north)}{\titlex}{\titley}
\draw[line width=1mm, dodger]($(logo.east)!0.5!(title.west)$) +(0,\subtitley) -- +(0,\titley);

\end{tikzpicture}
\vspace{3cm}

\sffamily\groupnumber

\sffamily
\large

{\sffamily\reportauthors}

\sffamily \grouptutor

\tikz[remember picture,overlay]\node[anchor=south,inner sep=0pt] at (current page.south) {\includegraphics[width=\paperwidth]{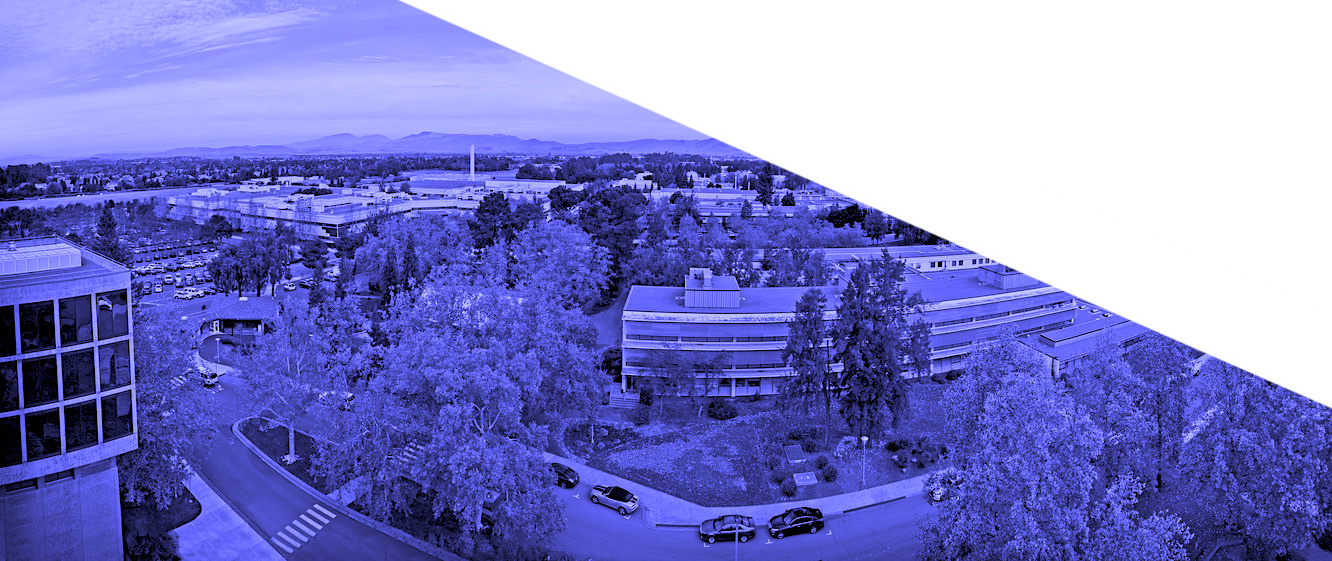}};

\mbox{}
\vfill
\sffamily \Large \textcolor{white}{\placeanddate} \\

\end{titlepage}

\newpage

\section*{Abstract}
\justifying
 { Although the dependence of convective core overshooting on mass has attracted much attention, no corresponding work exists for overshooting below a convective envelope.}
{We aim to quantify this relationship for pre-main sequence stars of intermediate mass ranging from $1.2 M_{\mathsf{sun}}$ to $6 M_{\mathsf{sun}}$.  These stars have a similar thermal and density structure, making this a suitable choice to isolate the effect of changing mass.}
{We produce a series of two-dimensional global simulations of stars using \music, a fully compressible, time-implicit hydrodynamics code.  
The stars that we select for this study are near the end of the pre-main sequence and are convectively unstable above 80\% of their stellar radius; they thus have a convective envelope that is shallower than the current sun.  For this series of stellar models, a simple scaling with luminosity, with a scaling exponent of $1/4$, accounts for the increasing overshooting with stellar mass.  This result has interesting similarities with the scaling found by \citet{baraffe2023study} for a range of intermediate mass and massive stars at the zero-age main sequence (ZAMS) that have convective cores. }

{\hypersetup{linkcolor=black} 
\tableofcontents\thispagestyle{empty}}
 \pagenumbering{arabic}

\vspace*{10mm}

\section{Introduction}

State of the art stellar evolution codes \citep{paxton2010modules,demarque2004y2,weiss2008garstec,siess2013binstar,christensen2008astec} evoke a diffusive process to model convective overshooting.  The diffusion coefficient \citep{freytag1996hydrodynamical,pratt2017extreme} used in the model can be set to change with the classification of the convective zone as non-burning, H-burning, He-burning, or metal-burning.  It can also be set to change at defined evolutionary points, such as the bottom of the asymptotic giant branch, or during the third dredge up \citep{herwig2000evolution,lugaro2003s}. Aside from these abrupt changes, the diffusion coefficient is typically locked to a percentage of the pressure scale height measured at the convective boundary.  However, there is no theoretical reason for convective overshooting, or indeed other convective properties, to change in the same way that the pressure scale height changes as a star evolves. 


The properties of convection and convective overshooting should be expected to change with a star's mass as well as the distribution of mass within a star.  This expectation stems from our fundamental understanding of thermal convection.   Convection is characterized by the Rayleigh number \citep[e.g.][]{featherstone2016spectral,he2012transition,van2012flow,chilla2012new} which quantifies the strength of the
upward motion of a fluid parcel due to positive buoyancy
relative to the strengths of drag and diffusion.
The Rayleigh number is linearly proportional to the gravitational acceleration $g$.  In the interior of a star, the gravitational acceleration is typically calculated from a Poisson equation for the gravitational potential, and described as \emph{self-gravity}; this results in a gravitational acceleration that changes as a function of a star's interior radius \citep{kippenhahn1990stellar}, and with the mass enclosed within that radius.    From experiments that isolate the impact of the magnitude and orientation of the gravitational acceleration, the gravitational acceleration changes basic properties of a convective flow  \citep[e.g.][]{egbers2003geoflow,futterer2013sheet,zaussinger2020rotating}.   It is thus reasonable to assume that, for a different stellar mass, the dynamics of stellar convection should change.  Understanding how the depth of convective overshooting changes with stellar mass, which is an input parameter for stellar evolution calculations, is directly useful in constructing overshooting models for stars.  It may also lead to better modeling of stars that are affected by mass loss \citep[e.g.][]{renzo2017systematic,higgins2019massive} or mass accretion \citep[e.g.][]{gibson2018mesa,goodwin2020binary}, situations that are of particular interest for binary systems. 

For overshooting above a convective core, several works \citep{umezu1995mass,ribas2000mass,vandenberg2006victoria,claret2016dependence,claret2018dependence,claret2019dependence,costa2019mixing} have explored the dependence on stellar mass by fitting stellar observations with the output of stellar evolution calculations.    Recent works \citep{lindsay2024fossil,viani2020examining}, have also estimated convective core overshooting using asteroseismology. These observational works have already shown that, for more massive stars, the overshooting depth above a core is larger relative to the pressure scale height; the assumption that overshooting is locked to the pressure scale height does not hold.  
  Measurements of the overshooting depth in multi-dimensional simulations of stars with core convection \citep{baraffe2023study} support a direct relationship between mass and overshooting, which is not accounted for by changes in the pressure scale height.  However, core convection is distinct from envelope convection in stars; core convection involves significantly higher density flows, slower velocities, and active feedback from chemical mixing.  
The dependence of convective envelope overshooting on mass therefore warrants specific study.

This work is structured as follows: 
 In Section \ref{sec:struc}, we discuss the ten 1D stellar structure models that we produce using the MESA code.
 In Section \ref{sec:sim} we introduce the numerical framework that we use to produce 2D hydrodynamic simulations for each of these stars using the \music code.  
 In Section \ref{secresultsid}, we discuss statistics related to the convective dynamics in our ten stellar simulations.  Using the extreme value theory (EVT) analysis of convective overshooting \citep{pratt2017extreme}, we relate the depth of overshooting beneath a convective envelope to the mass of the star. 
We discuss the filling factor for stellar convection and how it relates to mass and the overshooting depth. In Section \ref{secconc}, we discuss the implications of these results for future stellar evolution calculations.

\section{Stellar Structures \label{sec:struc}}

The structures for the pre-main sequence stars in the present study were produced by running the one-dimensional stellar evolution code Modules for Experiments in Stellar Astrophysics (MESA) \citep{paxton2010modules}.    The simple inlist provided in Appendix A was used in MESA version 10398.  For each evolutionary track, the only parameter that has been changed in this inlist is the initial mass; the mass has been varied from 1.2 to 6 solar masses, sampling 10 different values as described in Table~\ref{tab:mesaparams}.   Our evolutionary models do not include mass-loss or mass-accretion, so that each star has the same mass at every point along its evolutionary track.  

\begin{figure}
  \begin{minipage}[c]{0.6\textwidth}
  \resizebox{4.in}{!}{\includegraphics{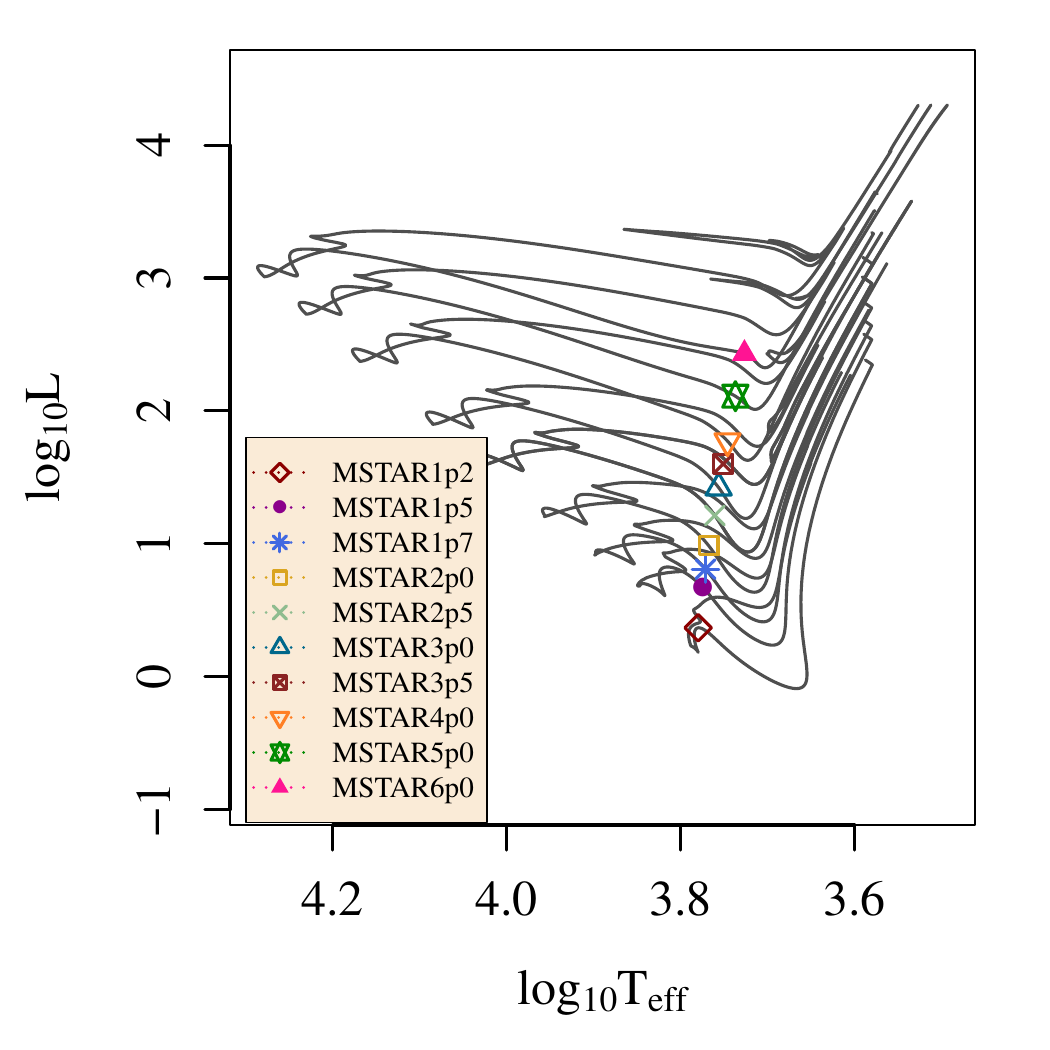}}
  \end{minipage}
  \hspace{5mm}\begin{minipage}[c]{0.3\textwidth}
\caption{Hertzsprung--Russell (HR) diagram showing the evolutionary paths of the ten stars produced using MESA.  The luminosity is expressed in units of solar luminosity. The points associated with each stellar structure that we use to perform hydrodynamic simulations is indicated by a colored symbol. \label{fig:hrdiagram}}
  \end{minipage}
\end{figure}

\begin{figure}
  \begin{minipage}[c]{0.8\textwidth}
  \resizebox{5.in}{!}{\includegraphics{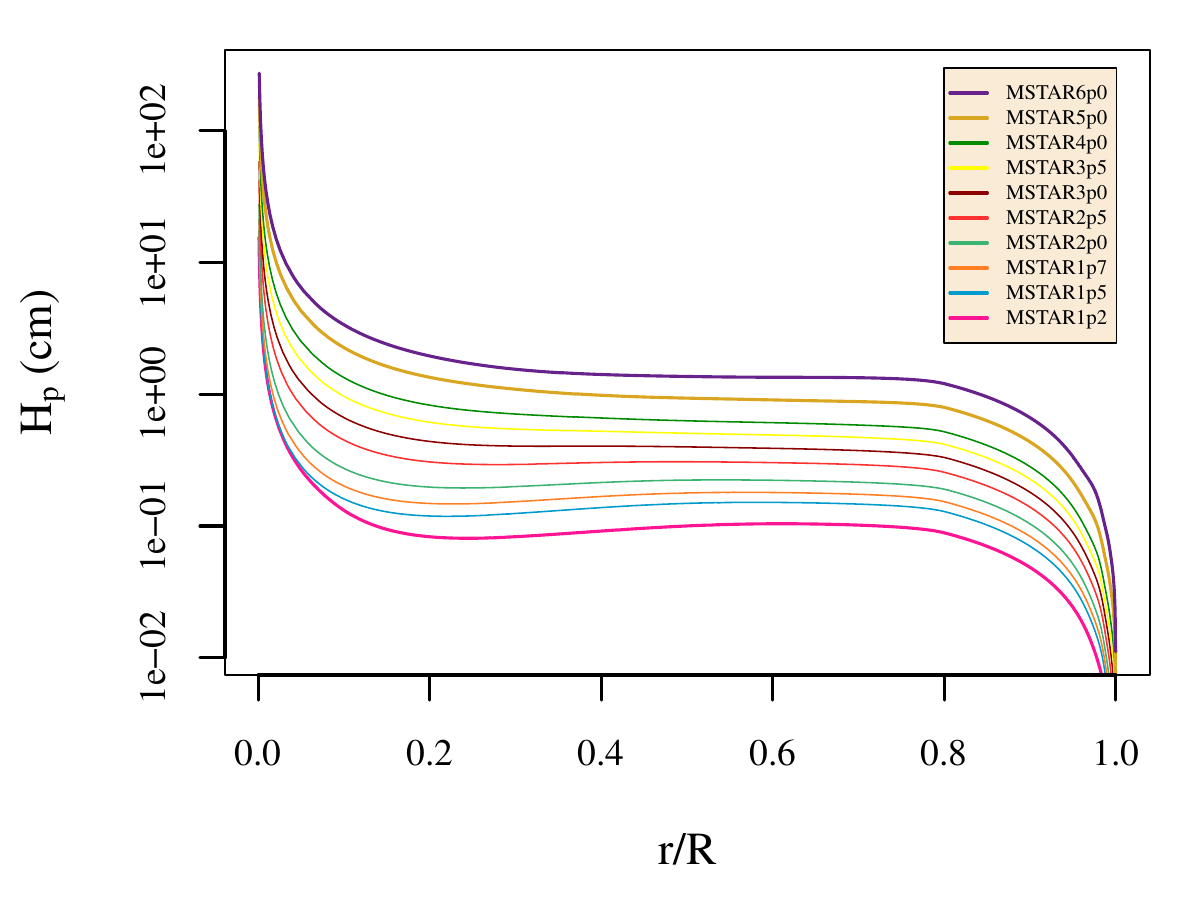}}
  \end{minipage}
  \hspace{5mm}\begin{minipage}[c]{0.15\textwidth}
\caption{Radial profile of the pressure scale height for the ten stars produced using MESA. \label{fig:hp}}
  \end{minipage}
\end{figure}

To formulate a study of convective overshooting, we select the first point along the evolutionary track where the star has an outer convective envelope with a convective boundary at $R_{\mathsf{CB}} = 0.8 R$, where $R$ is the radius of the star at the photosphere.  This allows us to 
compare convection zones with the same aspect ratio, an important consideration highlighted in fundamental studies of convection \citep[e.g.][]{bailon2010aspect,stevens2011effect,he2020aspect}.  The stars in our study are all located near the end of the pre-main-sequence phase of evolution; differences in their convection are not caused by differences in their evolutionary state. 
On the evolutionary tracks for these stars (Figure~\ref{fig:hrdiagram}), the stellar structure models used for our hydrodynamic simulations smoothly increase in luminosity as mass is increased.  The basic parameters of our ten stars are summarized in Table~\ref{tab:mesaparams}.  A comparison of the pressure scale height for these stars shows that the stellar structure smoothly increases with the star's mass, suggesting that this is a closely comparable series of stars (see Figure~\ref{fig:hp}).

\begin{table*}
\rowcolors{2}{gray!10}{white}
\centering
\caption{Parameters of  stellar structure models produced with MESA.  
 The stellar mass $M$ is given in units of the solar mass; the age is provided in years.  These models are selected so that the radial position of the convective boundary, as determined by the Schwarzschild criterion, is at $r_{\mathsf{CB}}/R = 0.8$, and the radius of the star at the photosphere $R$ is provided. The stellar luminosity at the photosphere $L$ is given in units of the solar luminosity.  The effective temperature at the photosphere in Kelvin, and log of the acceleration of gravity at the photosphere are also provided.   Each model has metallicity $z=0.02$ and helium mass fraction $y = 0.28$.
}
\label{tab:mesaparams}
\begin{tabular}{l|cccccc} 
\hline
{Name} & {$M/M_{\mathsf{sun}}$}&  {Age  (1e5 yrs)} & {$R$ ($10^{12}$cm)} & {$L/L_{\mathsf{sun}}$} &  {$T_{\mathsf{eff}} (10^3$K)}&  {$\log_{10} g_{\mathsf{eff}}$  (cm/s$^2$)} \\
\hline
MSTAR1p2 & 1.2   &    208.8 & $0.098$    &  2.33     & 6.02  & 4.222 \\ 
MSTAR1p5 & 1.5   &    106.8 & $0.142$    &  4.70     & 5.95  &  3.993 \\ 
MSTAR1p7 & 1.7   &    75.3   & $0.169$    &  6.45     & 5.91  &  3.898 \\ 
MSTAR2p0 & 2.0   &    48.2  & $0.211$    &  9.70     & 5.86  &  3.777 \\ 
MSTAR2p5 & 2.5   &    25.6   & $0.282$    &  16.35   & 5.76  &  3.619 \\ 
MSTAR3p0 & 3.0   &    15.1   & $0.367$  & 26.45    & 5.70   &  3.471  \\ 
MSTAR3p5 & 3.5   &    9.26 & $0.461$  & 39.90    & 5.64   &  3.338  \\ 
MSTAR4p0 & 4.0   &    5.87 & $0.575$   & 59.28  & 5.57   &  3.205 \\ 
MSTAR5p0 & 5.0   &    2.55 & $0.885$  & 129.06  & 5.46   &  2.928 \\ 
MSTAR6p0 & 6.0   &    1.22 & $1.340$  & 269.98  & 5.33   &  2.645 \\ 
\hline
\end{tabular}
\end{table*}

\begin{figure}
  \begin{minipage}[c]{0.6\textwidth}
  \resizebox{4.in}{!}{\includegraphics{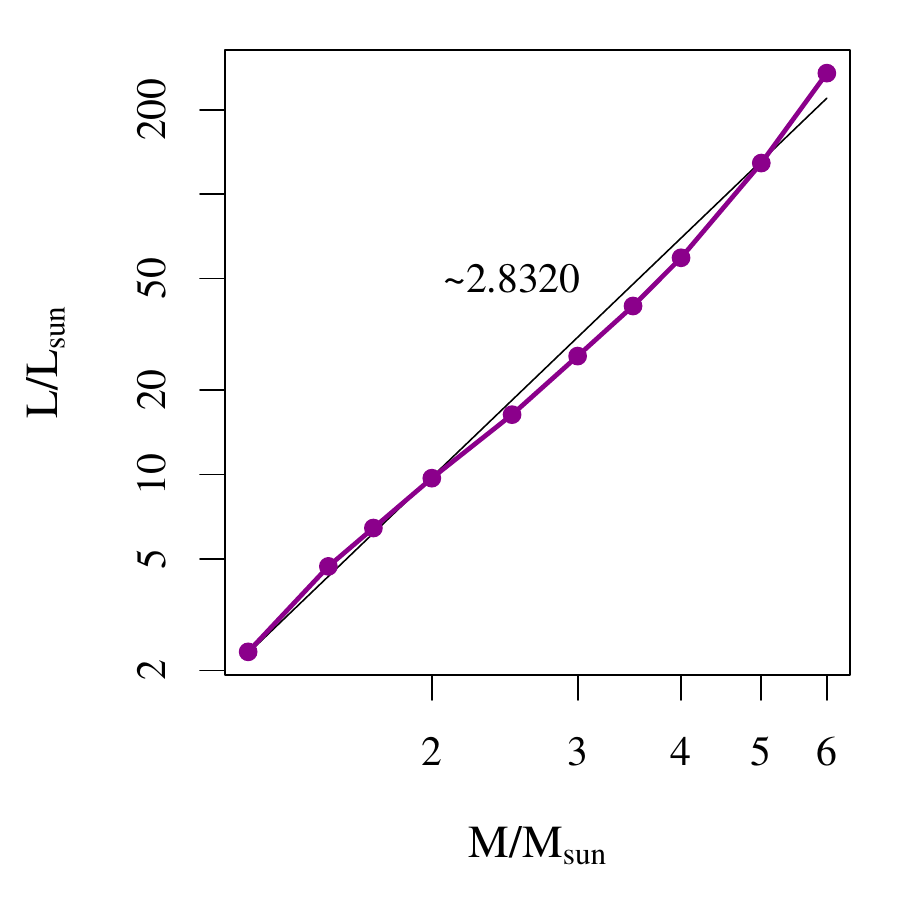}}
  \end{minipage}
  \hspace{5mm}\begin{minipage}[c]{0.3\textwidth}
\caption{Surface luminosity vs mass for the stellar structures described in Table~\ref{tab:mesaparams}.  Both of these quantities are given in solar units.  A power-law fit  provides a reasonable fit to the series of MESA stellar structures that we produce.   \label{fig:lummass}}
  \end{minipage}
\end{figure}

To understand how the structures of these stars change with mass, we first examine the mass--luminosity relation (see Figure~\ref{fig:lummass}).  For main-sequence stars, we would expect the scaling exponent relating luminosity with mass to be in the range of 3.5 -- 4.0 for these masses \citep{torres1997hyades,duric2004advanced,cuntz2018mass}.  Because these are pre-main-sequence stars, there is no observational scaling for the mass--luminosity relation.  For the present range of masses, we nevertheless employ a linear regression fit to find that this scaling exponent is approximately 2.8.  Although this fit is not equally good in all regions of this mass range, it does describe the overall increase in luminosity with mass fairly well for these stars that are approaching the ZAMS.

\begin{figure}
  \resizebox{3.3in}{!}{\includegraphics{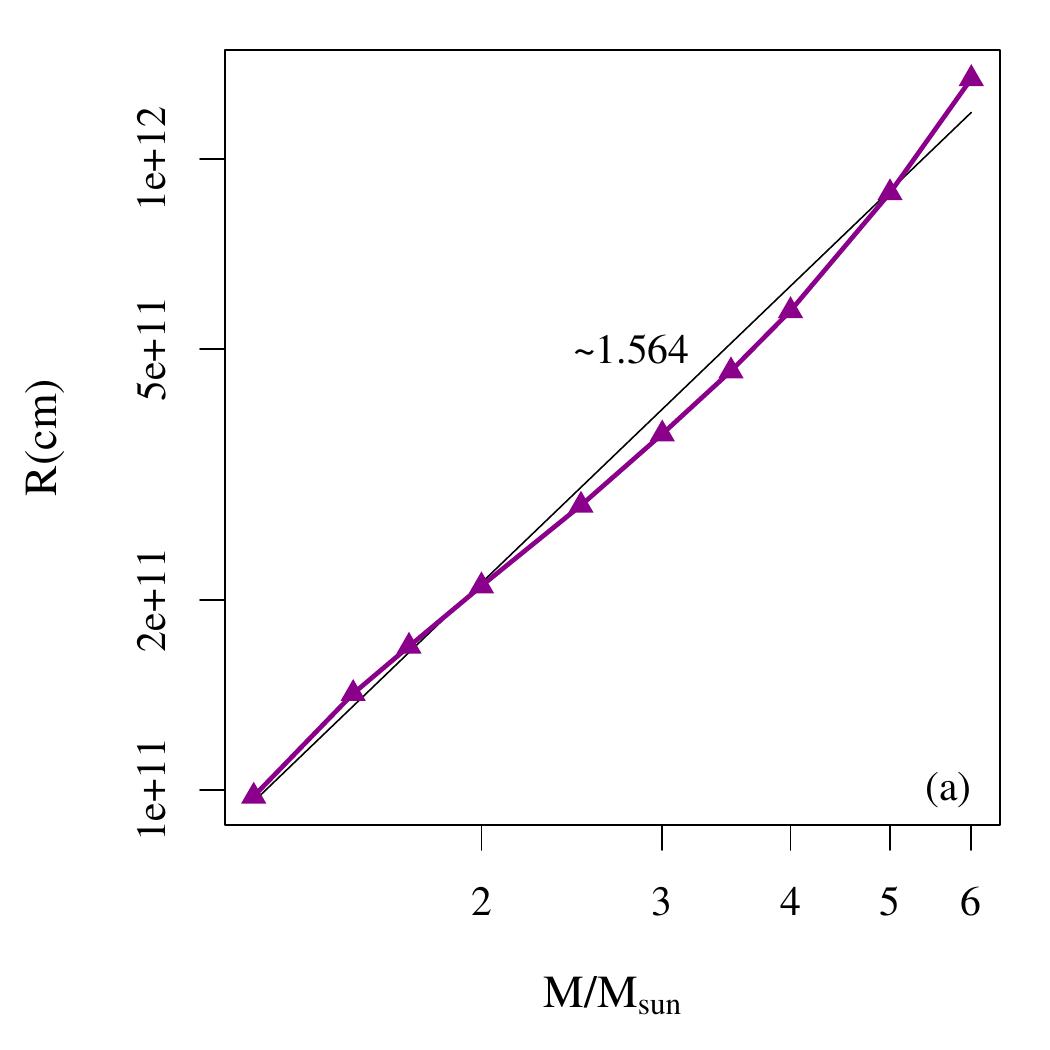}}\resizebox{3.3in}{!}{\includegraphics{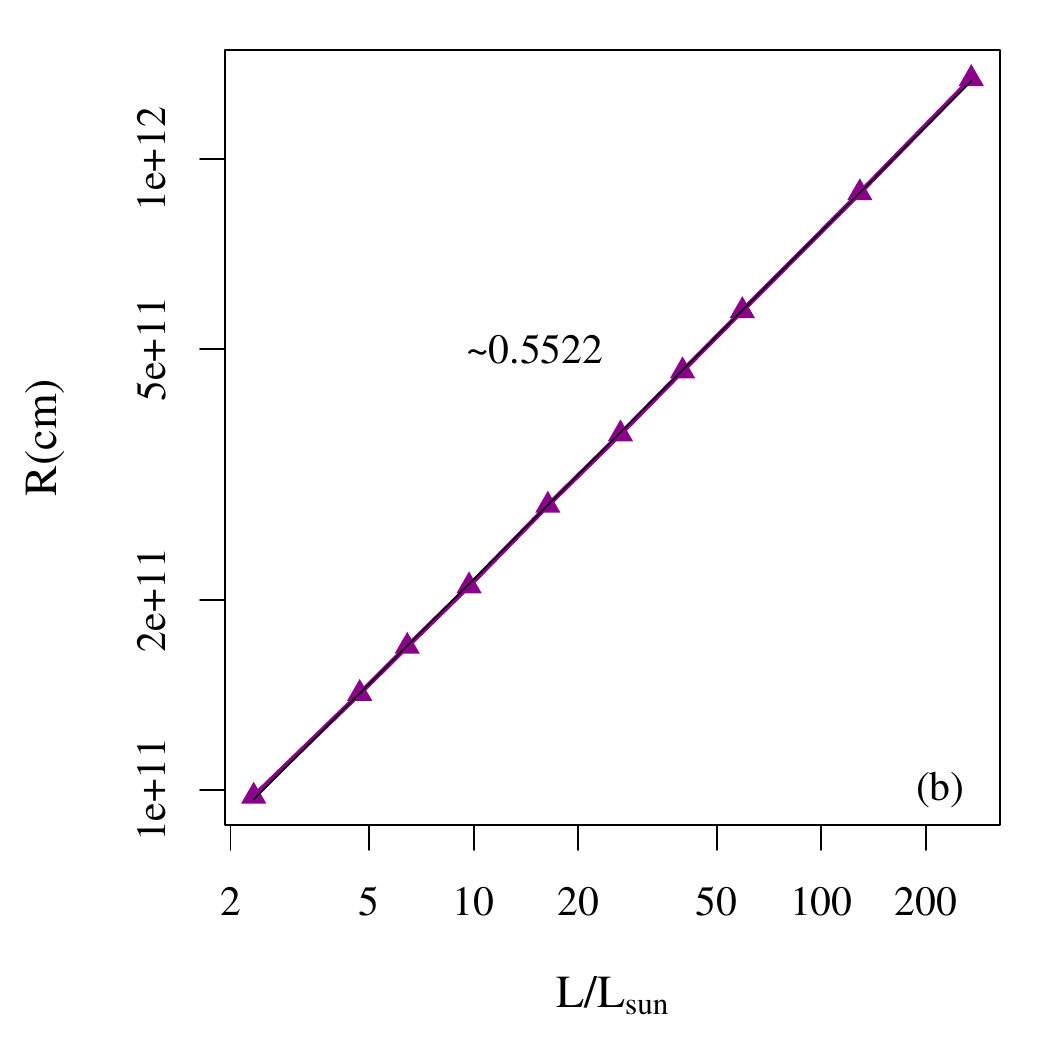}}
\caption{Radius of the photosphere $R$ in centimeters (a) vs mass, and (b) vs luminosity in solar units for the stellar structures described in Table~\ref{tab:mesaparams}.  Linear regression fits are shown for each log-log plot.   \label{fig:radmass}}
\end{figure}

We also examine how the radius of the photosphere, also simply described as the stellar radius $R$, changes with mass for these stellar structures (see Figure~\ref{fig:radmass}).  Since observers often plot mass-radius relations for stars at a fixed age, a direct comparison to an observationally calculated scaling is not possible for the series of stars we have selected for our study.   A linear regression fit produces a scaling relationship between the mass and radius 
\begin{eqnarray}\label{rscalemass}
R \sim \left( \frac{M}{M_{\mathsf{sun}}} \right)^{1.564}~.
\end{eqnarray}
The fit here is of a similar quality to the mass--luminosity relation, and shows that, for stars at a similar stage in evolution, the stellar radius increases smoothly as its mass increases.  The lowest and highest masses fit this scaling law as well as the other masses.  The stellar radius of these stars also has a clear scaling with luminosity.

\begin{figure}
  \resizebox{3.3in}{!}{\includegraphics{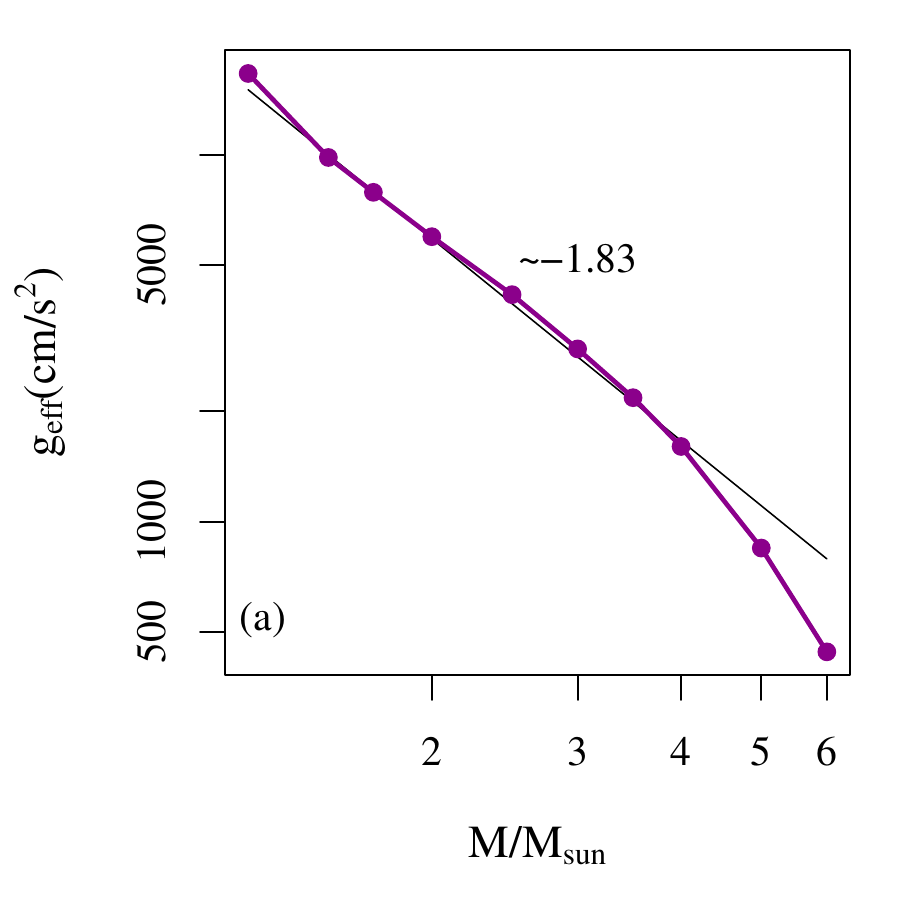}}\resizebox{3.3in}{!}{\includegraphics{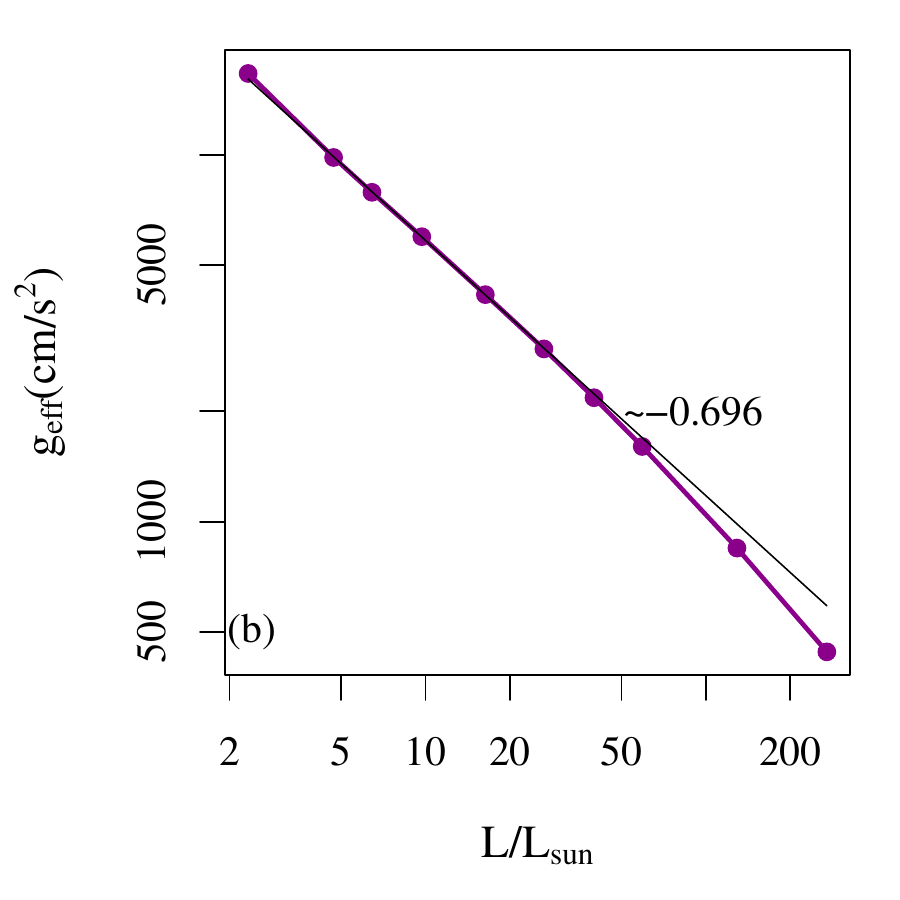}}
\caption{Gravitational acceleration at the photosphere $g_{\mathsf{eff}}$  (a) vs mass, and (b) vs luminosity for the stellar structures described in Table~\ref{tab:mesaparams}.  Linear regression fits are shown for each log-log plot.   \label{fig:loggmass}}
\end{figure}

For stars in this series, the gravitational acceleration at the photosphere is smaller as we examine higher mass (see Figure~\ref{fig:loggmass}).  This is a natural consequence of the radius also increasing, since $g \propto M/R^2$.  In this case a linear regression fit shows that gravitational acceleration is decreasing with a scaling exponent of $1.83$ with increasing mass.  The 5 and 6 solar mass stars depart from this scaling law for gravity more strongly than the other masses, and more strongly than they did from the luminosity or radius trends.  This indicates that differences are beginning to appear in the stellar structures for these higher masses.  Because this deviation is not yet severe, we still include these masses in our study.  However the differences in gravitational acceleration may lead to other differences in the properties of convection.

\section{Simulations \label{sec:sim}}

For the series of ten stellar structures of pre-main-sequence stars described in Table~\ref{tab:mesaparams}, we perform 2D implicit large eddy simulations (ILES) using the MUltidimensional Stellar Implicit Code (\music) \citep[as described in][]{goffrey2017benchmarking,prattspherical}. 
We study the dynamics of convection, and particularly convective boundary mixing, in these MESA stellar structures.  
Our simulations only take convection into account; the possibility of studying additional physical effects such as rotation, a tachocline, chemical mixing, and magnetic fields are omitted from the current study, which focuses on expanding our understanding of convection and overshooting in a stratification that is realistic to stars. 
The \music code solves the inviscid compressible hydrodynamic equations for density $\rho$, momentum $\rho \vec{u}$, and internal energy $\rho e$:
\begin{eqnarray} \label{densityeq}
\frac{\partial}{\partial t} \rho &=& -\nabla \cdot (\rho \vec{u})~,
\\ \label{momeq}
\frac{\partial}{\partial t} \rho \vec{u} &=& -\nabla \cdot (\rho \vec{u} \vec{u}) - \nabla p + \rho \vec{g} ~,
\\ \label{ieneq}
\frac{\partial}{\partial t} \rho e &=& -\nabla \cdot (\rho e\vec{u}) +p \nabla \cdot \vec{u} + \nabla \cdot (\chi \nabla T) ~.
\end{eqnarray}
\music solves these equations using a finite volume method, a MUSCL method \citep{thornber2008improved} of interpolation, and a van Leer flux limiter \citep[as described in][]{van1974towards,roe1986characteristic,leveque2006computational}.   For these 2D simulations, the finite volume method assumes azimuthal symmetry.  Time integration in the \music code is fully implicit, and uses a Jacobian free Newton-Krylov (JFNK) solver \citep{knoll2004jacobian} along with a  preconditioning method based on the solution of the physical equations \citep{maximepaper}.  The \music code uses an adaptive time step, which is constrained identically for all simulations studied in this work.   In eq.~\eqref{momeq}, the gravitational acceleration $\vec{g}$ is a radial vector consistent with the stellar structures produced by MESA, and not evolved by our simulations.  

In eq.~\eqref{ieneq}, the thermal conductivity $\chi = 16 \sigma T^3/3 \kappa \rho$ is defined using the Stefan-Boltzmann constant $\sigma$ and the Rosseland mean opacity $\kappa$.  The thermal conductivity is thus realistic to the pre-main-sequence stars studied here, and has not been enhanced.
The compressible hydrodynamic equations \eqref{densityeq}-\eqref{ieneq} are closed by determining the gas pressure $p(\rho,e)$ and temperature $T(\rho,e)$ from a tabulated equation of state
constructed from the 2005 update of the OPAL EOS tables of  \citet{rogers2002updated}.
The Rosseland mean opacities are interpolated from the OPAL
tables \citep{iglesias1993radiative,iglesias1996updated}  for higher temperatures in the range $3.75 \leq \log_{10} T \leq 8.7$ and from  \citet{ferguson2005low} for lower temperatures above $ \log_{10} T > 2.7$.  These tabulated equations of state and opacities are identical to those used to produce the stellar structures in MESA. 
    
\begin{figure*}
\includegraphics[height=.65\columnwidth]{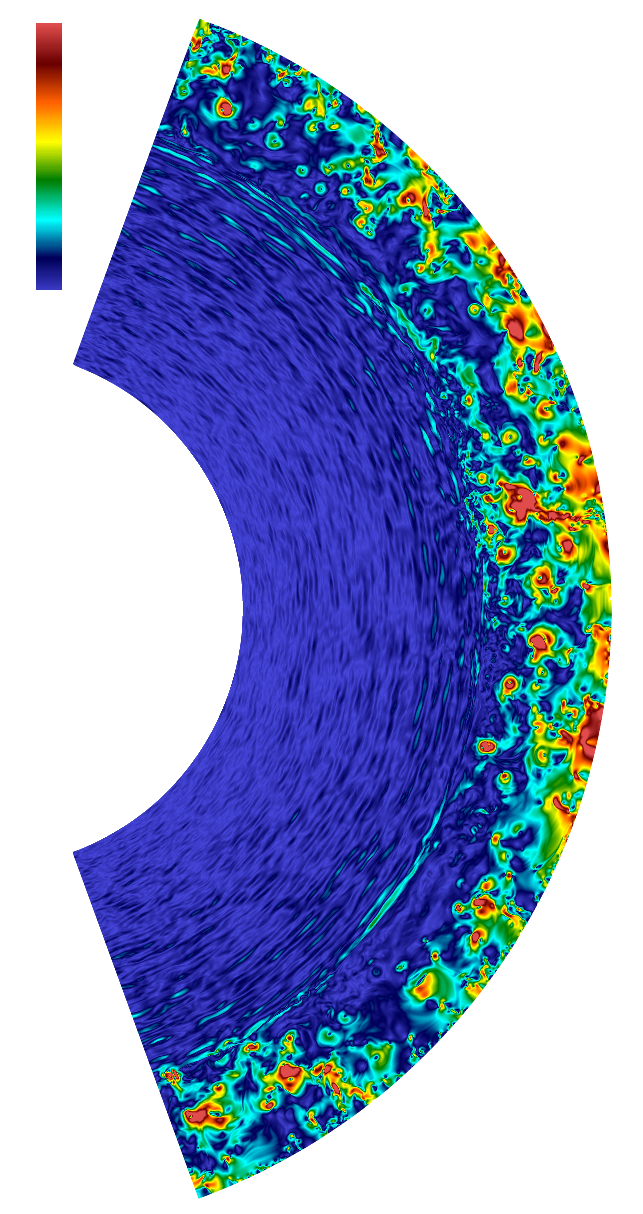}\includegraphics[height=.65\columnwidth]{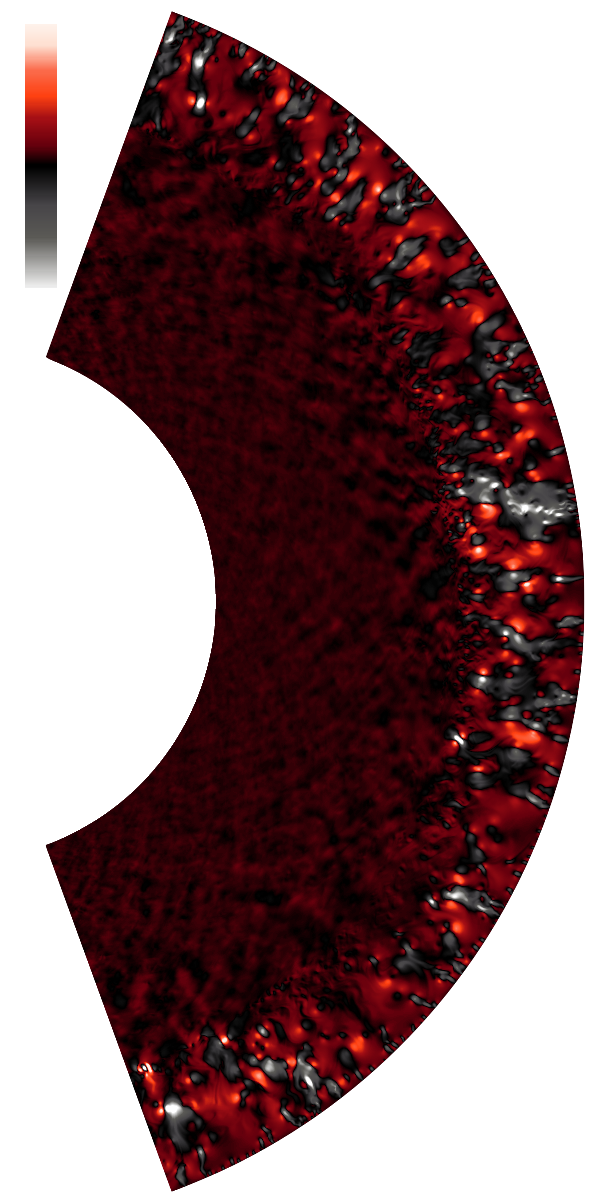}\includegraphics[height=.65\columnwidth]{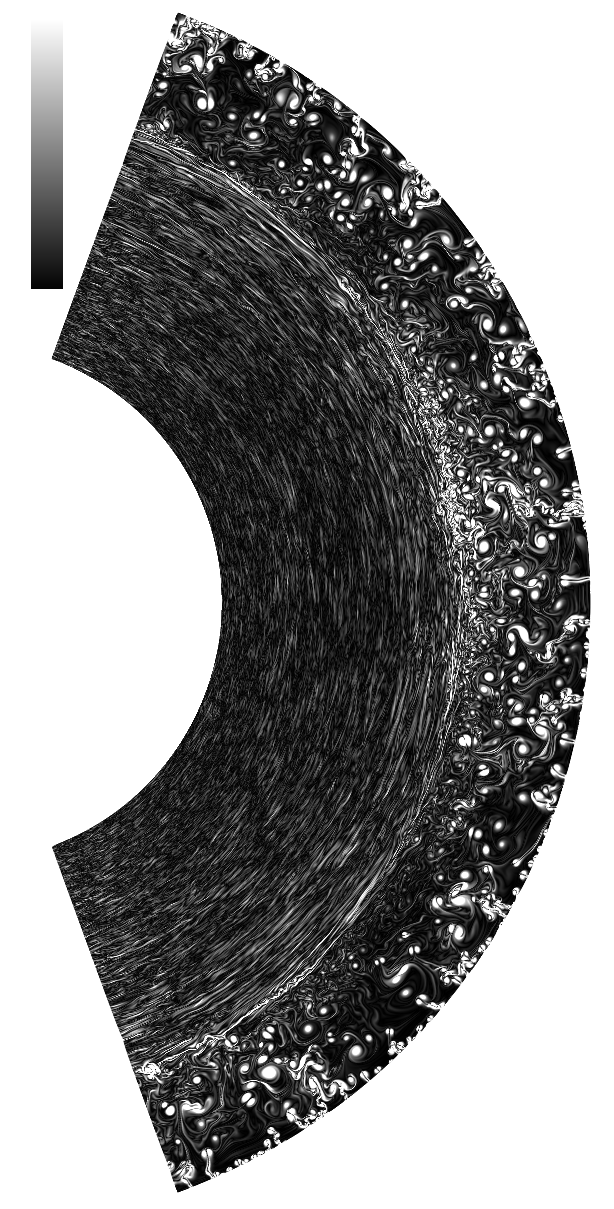}
\caption{Visualizations from simulation MSTAR5p0: (left) a typical snapshot of the velocity magnitude, (middle) radial velocity, and (right) vorticity magnitude.  The color scales for velocity and vorticity magnitude are set so that the minimum color is zero.  The color scale for the radial velocity is set so that black (directly in the middle of the color scale) indicates zero, grey indicates inward flows, and red indicates outward flows.  \label{fig:viz5}}
\end{figure*}

\music solves the compressible hydrodynamic equations \eqref{densityeq}-\eqref{ieneq} in a spherical shell using spherical coordinates in 2D:  radius $r$, and angle $\theta$.   Ten \music simulations are summarized in Table~\ref{tab:musicparams}.  Each simulation volume begins at $20 \degree$ from the north pole, and ends $20 \degree$ before the south pole.  Each simulation has an inner radius of $R_{\mathsf{i}}/R=0.4$ and an outer radius of $R_{\mathsf{o}}/R=0.97$.  Our radial grid is determined so that the pressure scale height at the convective boundary $H_{p\mathsf{CB}}$ is resolved with approximately the same number of grid cells in every simulation.  This resolution is $H_{p\mathsf{CB}}/\Delta r \approx 240 \pm 10$, a high resolution compared to previous works \citep[e.g.][]{pratt2017extreme}.  The simulation MSTAR1p2 has a grid of $r \times \theta = 2224 \times 2224$.   Because the convection zone is the same size in each star, and because the simulations are produced in volumes that are identically shaped, the different stars are visually indistinguishable.  A visualization of the velocity magnitude, radial velocity, and vorticity magnitude from simulation MSTAR5p0 is provided in Figure~\ref{fig:viz5}.  

\begin{table*}
\rowcolors{2}{gray!10}{white}
\centering
\caption{Parameters describing 2D hydrodynamic simulations performed with the \music code.   The inner radius of the simulation $R_{\mathsf{i}}$, the radius of the convective boundary $R_{\mathsf{CB}}$,
the radius of the outer boundary of the simulation  $R_{\mathsf{o}}$ are provided in units of the stellar radius.  The time-average global convective turnover time $\tau_{\mathsf{conv}}$ is computed as described in eq.~\eqref{eqtauconv}, and an error estimate is provided in the form of its standard deviation in time over the period of steady-state convection.  The total simulation time in units of the average convective turnover time is also provided.  We measure the resolution at the convective boundary using a non-dimensional quantity constructed from the pressure scale height at the convective boundary, and the size of our radial grid space $H_{p\mathsf{CB}}/\Delta r$.  \label{tab:musicparams}}
\begin{tabular}{|l|cccc|} 
\hline
 {Name} & ($R_{\mathsf{i}}$, $R_{\mathsf{CB}}$,$R_{\mathsf{o}}$)$/R$  & {$\tau_{\mathsf{conv}}(10^5$s)} & {time ($\tau_{\mathsf{conv}}$)} 
& {$H_{p\mathsf{CB}}/\Delta r$}  \\
\hline
MSTAR1p2 & (0.4, 0.8, 0.97) &  $1.65 \pm 0.14$ & 59   &  241 \\ 
MSTAR1p5 & (0.4, 0.8, 0.97) &  $1.73 \pm 0.12 $ & 60   &  243 \\ 
MSTAR1p7 & (0.4, 0.8, 0.97) &  $1.91 \pm 0.11 $ & 61   &  244 \\ 
MSTAR2p0 & (0.4, 0.8, 0.97) &  $2.11 \pm 0.15$ & 190 &  243 \\ 
MSTAR2p5 & (0.4, 0.8, 0.97) &  $2.48 \pm 0.10$ & 60   &  245 \\ 
MSTAR3p0 & (0.4, 0.8, 0.97) &  $2.67 \pm 0.12$ & 69   &  235  \\ 
MSTAR3p5 & (0.4, 0.8, 0.97) &  $2.96 \pm 0.24$ & 60   &  245 \\ 
MSTAR4p0 & (0.4, 0.8, 0.97) &  $3.12 \pm 0.16$ & 49   &  247 \\ 
MSTAR5p0 & (0.4, 0.8, 0.97) &  $3.25 \pm 0.19$ & 51   &  242 \\ 
MSTAR6p0 & (0.4, 0.8, 0.97) &  $3.25 \pm 0.21$ & 34   &  245  \\ 
\hline
\end{tabular}
\end{table*}


In \citet{prattspherical,pratt2020comparison}, we studied the placement of boundaries and the choice of boundary conditions; we found that these choices do affect the physical outcome of our hydrodynamic simulations.  For the present series of simulations, we hold the energy flux and luminosity constant on the outer radial boundary, at values established from the MESA stellar structure.  All of the simulations include up to 0.97 of the stellar radius at the photosphere, where this boundary condition is applied.  We impose periodicity on all physical quantities at the boundaries in $\theta$.   In velocity, we impose non-penetrative and stress-free boundary conditions in the radial directions.  The energy flux and luminosity are held constant at the inner radial boundary, at the value of the energy flux in the MESA stellar structure.   
At the outer radial boundary, we apply a hydrostatic equilibrium boundary condition on the density that maintains hydrostatic equilibrium by assuming a constant derivative of internal energy and constant radial acceleration due to gravity in the boundary cells \citep{hsegrimm2015realistic}.  
 On the inner radial boundary of the spherical shell, we impose a constant radial derivative on the density on all simulations except MSTAR3p0; simulation MSTAR3p0 uses a boundary condition on the density that maintains hydrostatic equilibrium by assuming constant internal energy at the inner radial boundary: we found that this better preserved the stratification of the original stellar structure.
These boundary conditions allow us to maintain the stratification of density at the boundaries of our simulation to the MESA structure of each star.

\section{Results \label{secresultsid}}

\subsection{A statistical picture of how convection changes over this mass range}

All results presented here are produced during a statistically stationary steady-state of stellar convection, a period where the time-averaged value of the total kinetic energy is well defined and not changing in time.  The time span of the simulation that the convection is in steady-state is noted in Table~\ref{tab:musicparams}.  Simulation MSTAR2p0 was run longer than the other simulations:  this was simply the first simulation that we performed in this series, the model for the rest of the simulations. 
To understand how the stellar mass can change the basic statistics of convection, we first calculate the convective turnover times.

The convective turnover time is a fundamental parameter produced in one-dimensional stellar evolution studies.
Following \citet{prattspherical}, we define the global time-scale $\tau_{\mathrm{conv}}$
\begin{eqnarray} \nonumber
\tau_{\mathrm{conv}} &=&  \Big\langle \int_{R_{\mathsf{CB}}}^{R_{\mathsf{o}}} \int_{\theta_2}^{\theta_1}  d \mathsf{V(r,\theta)}~ H_p(r,\theta,t)/ |u(r,\theta,t)|  ~/~ 
\\  \label{eqtauconv}
&~& \int_{R_{\mathsf{CB}}}^{R_{\mathsf{o}}} \int_{\theta_2}^{\theta_1}  d \mathsf{V(r,\theta)} \Big\rangle_t
\end{eqnarray}
In this expression, $H_p$ is the pressure scale height as a function of position and time, and $|u|$ is the velocity magnitude.  
The integration is volume-weighted and $d\mathsf{V(r,\theta)}$ is a volume element.  The integration covers the convection zone, between  $R_{\mathsf{CB}}/R=0.8$ and $R_{\mathsf{o}}/R=0.97$, and the full angular extent of our spherical shells.  The brackets $\langle ...\rangle_t $ indicate a time average, taken over the full time span of the simulation in steady state, noted in Table~\ref{tab:musicparams}.   The standard deviation in time is used to produce an error on $\tau_{\mathrm{conv}}$.  

\begin{figure}
\includegraphics[width=.5\columnwidth]{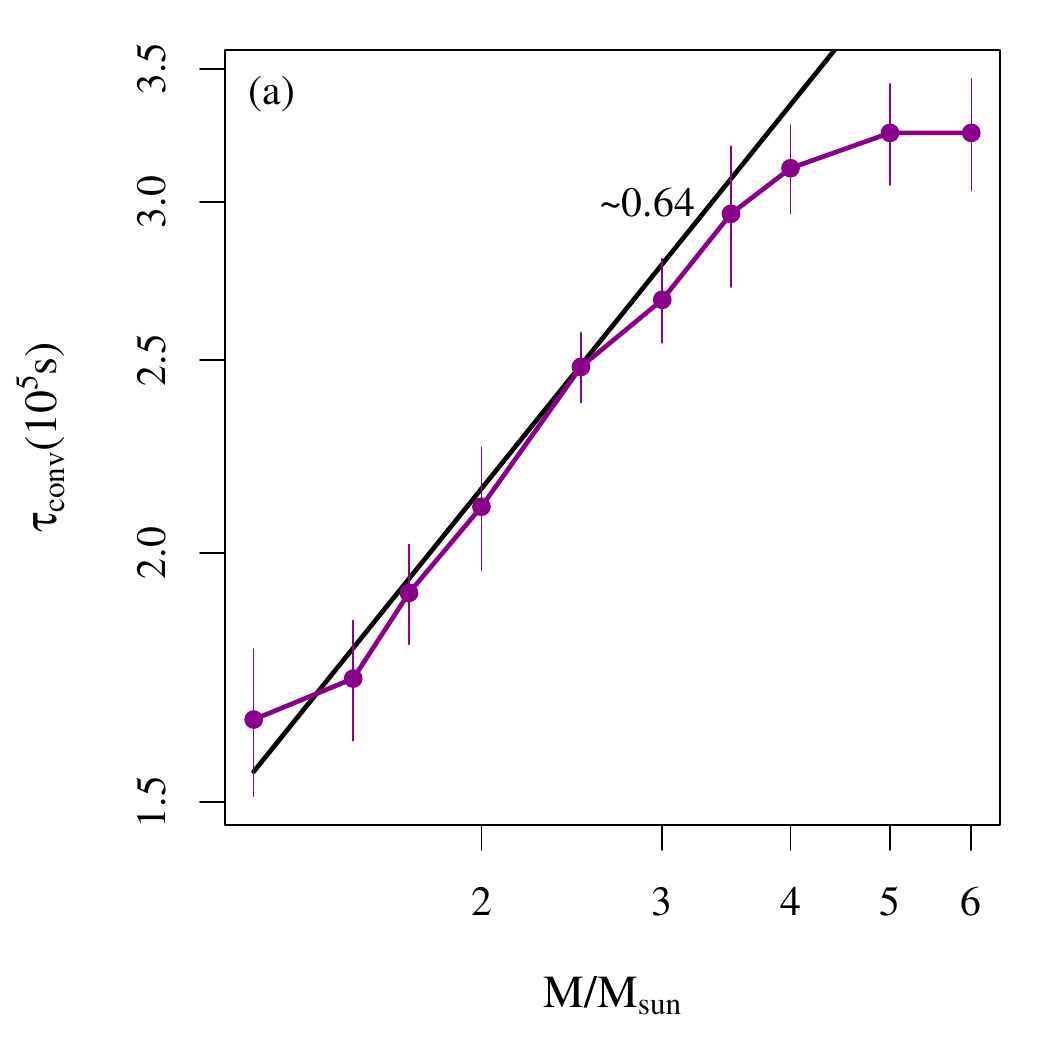}\includegraphics[width=.5\columnwidth]{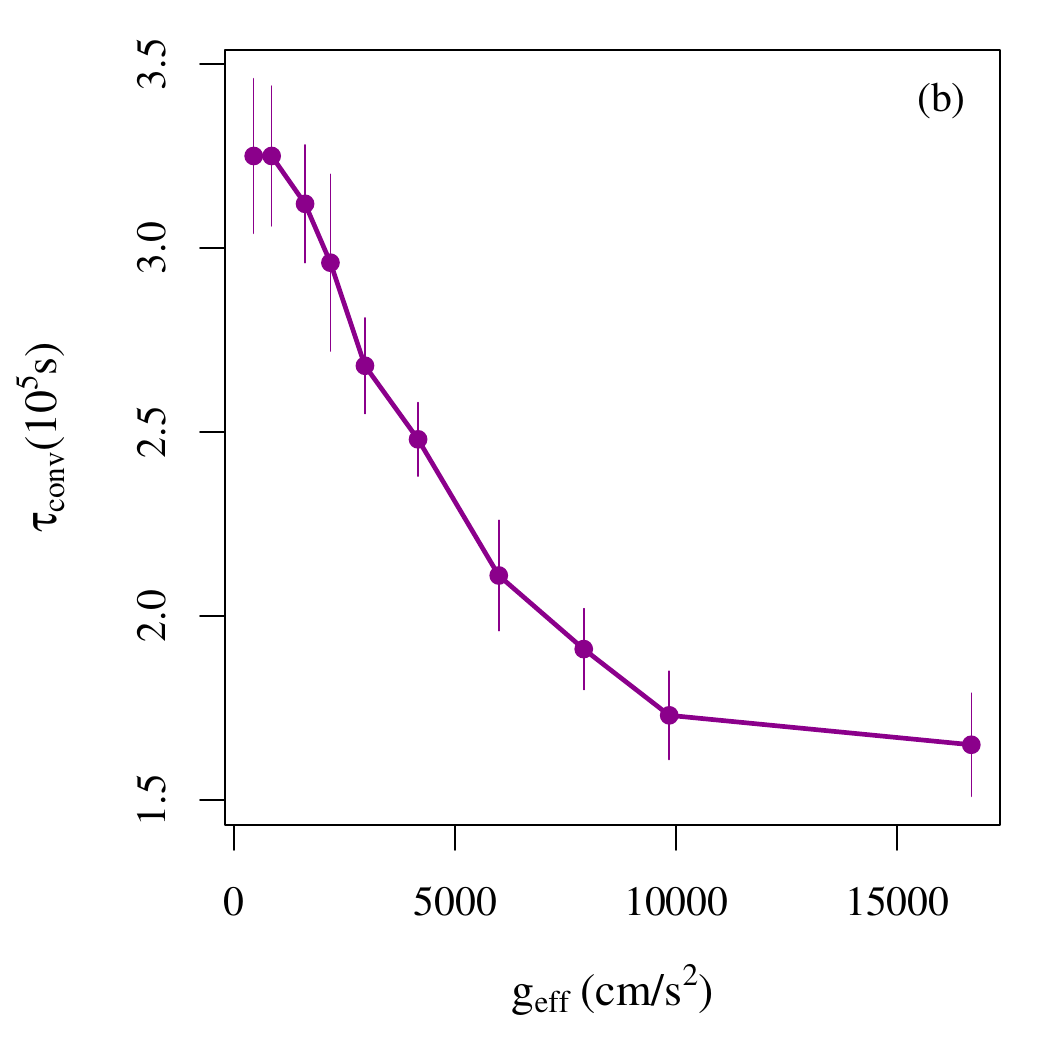}
\includegraphics[width=.5\columnwidth]{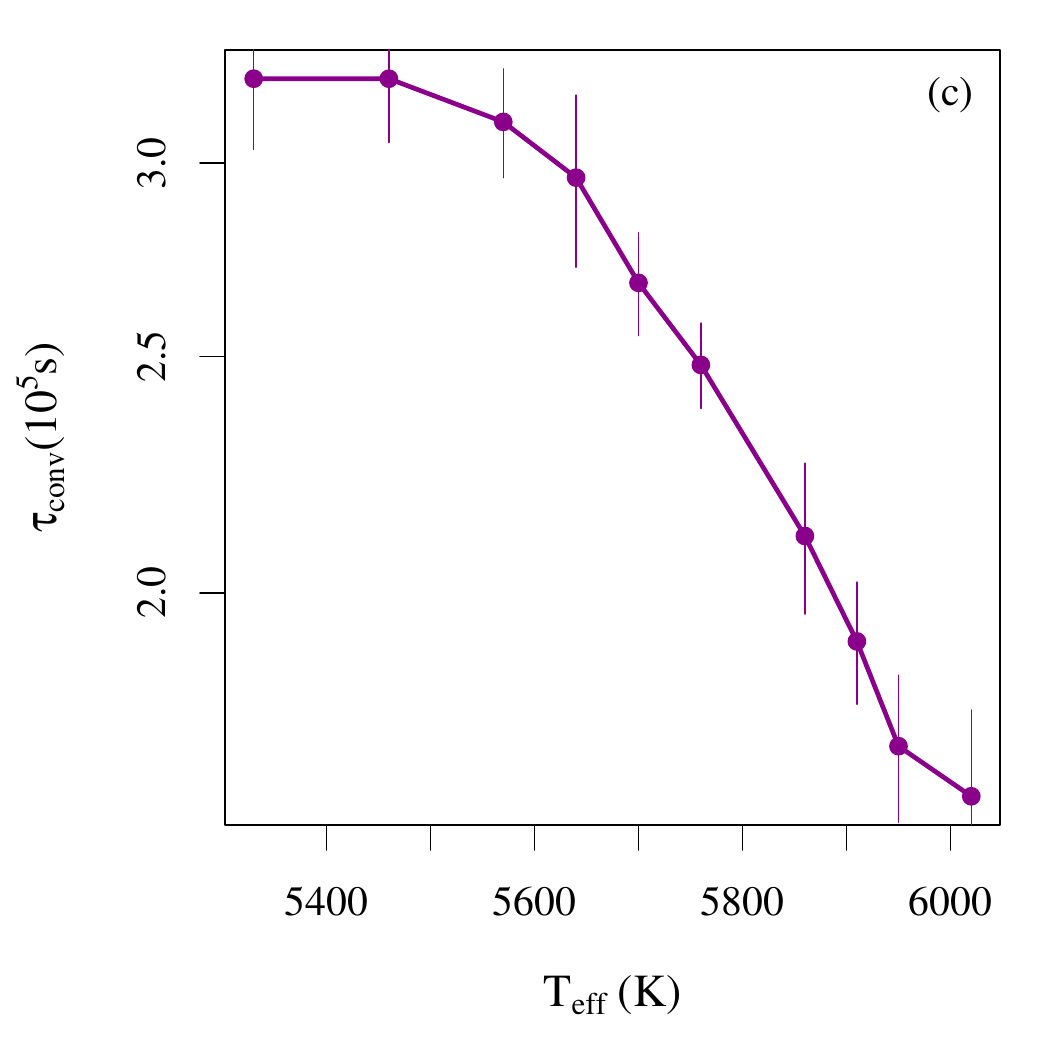}\includegraphics[width=.5\columnwidth]{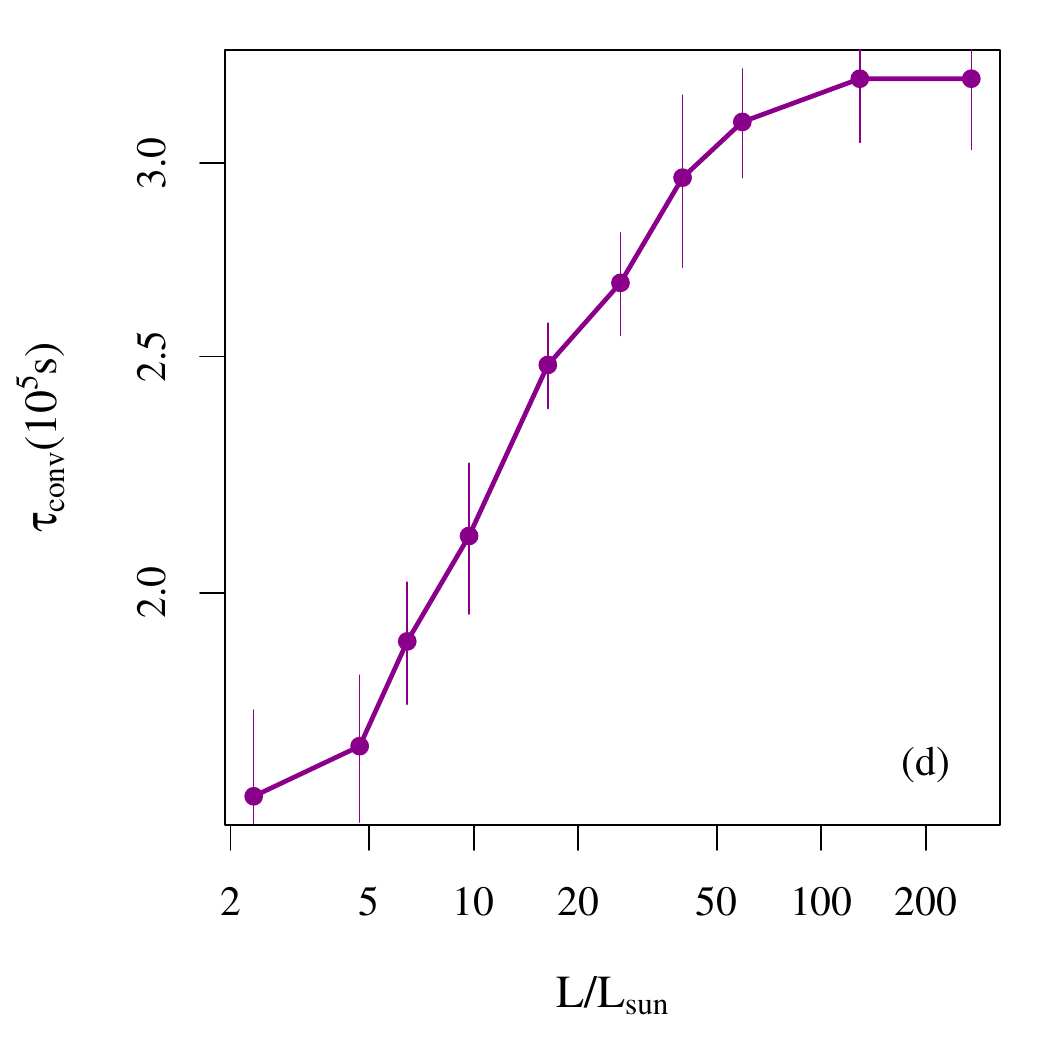}
\caption{Convective turnover time $\tau_{\mathsf{conv}}$ for the ten simulations in Table~\ref{tab:musicparams}: (a) trend with mass, (b) trend with gravitational acceleration at the photosphere, (c) trend with $T_{\mathsf{eff}}$, and (d) trend with luminosity.  Error bars are plotted based on the standard deviation in time for this time-averaged value. \label{fig:tau}}
\end{figure}

The convective turnover time exhibits a smooth monotonic increase with larger mass (see Figure~\ref{fig:tau}(a)).   In the range $2 \leq M/M_{\mathsf{sun}} \leq 4$ the convective turnover time increases rapidly with mass, following an approximate power-law scaling; a linear regression fit calculates the scaling exponent in this range to be
\begin{eqnarray}\label{tauscaling}
\tau_{\mathrm{conv}} \sim \left( \frac{M}{M_{\mathsf{sun}}} \right)^{0.64}~.
\end{eqnarray}
A small deviation from this scaling is seen for the 1.2 solar mass star; a larger deviation is evident for the stars that are 5 and 6 solar masses.   
To estimate the convective turnover time from the convective velocity $v_{\mathrm{conv}}$, astronomers sometimes use a simple dimensional scaling  $\tau_{\mathrm{conv}} \sim R_{\mathsf{ext}}/v_{\mathrm{conv}}$.  For the present study, the radial extent of the convective envelope for the present set of stars is $R_{\mathsf{ext}} = 0.2 R$. Using the scaling for velocity with luminosity $v_{\mathrm{conv}} \sim L^{1/3}$, and the scaling of $R$ with mass calculated from the stellar structures in eq.~\eqref{rscalemass}, along with the mass--luminosity relation that we found for these stars, this simple scaling would predict $\tau_{\mathrm{conv}} \sim M^{0.62}$.  This dimensional prediction results in a scaling exponent that is very similar to the one in  eq.~\eqref{tauscaling}, which was produced from the analysis of our hydrodynamic simulation data over long times.

The convective turnover time also decreases rapidly with increasing gravitational acceleration at the photosphere $g_{\mathsf{eff}}$ (see Figure~\ref{fig:tau}(b)) and with effective temperature $T_{\mathsf{eff}}$ (see Figure~\ref{fig:tau}(c)); it increases with luminosity with a trend steeper than mass (see Figure~\ref{fig:tau}(d)).  The data therefore indicates that no scaling exists for this full mass range with any of these fundamental parameters.   The values of the convective turnover time that we calculate for our series of pre-main sequence stars are a factor of 10 lower than for the stars on the ZAMS examined by \citet{cranmer2011testing}. 

\begin{figure}
\includegraphics[width=.5\columnwidth]{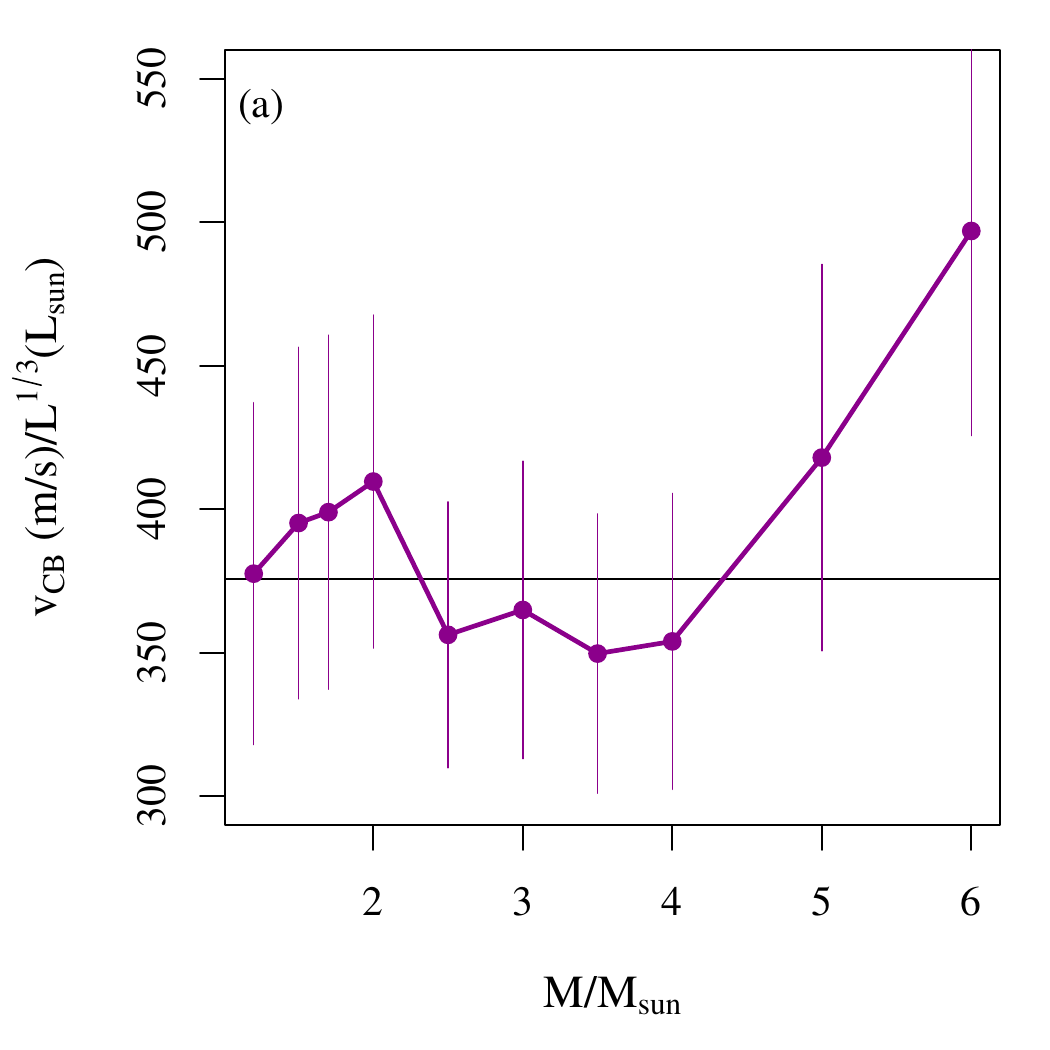}\includegraphics[width=.5\columnwidth]{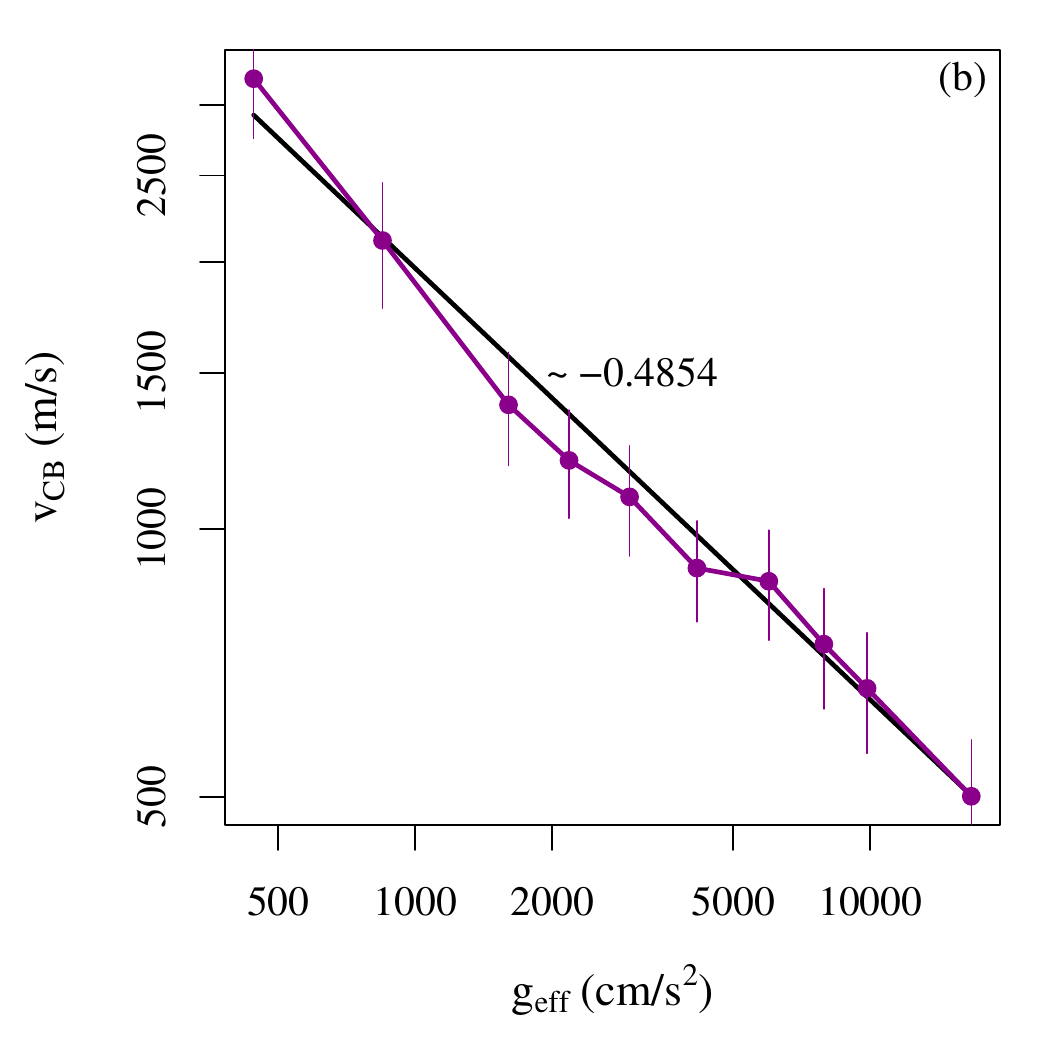}
\includegraphics[width=.5\columnwidth]{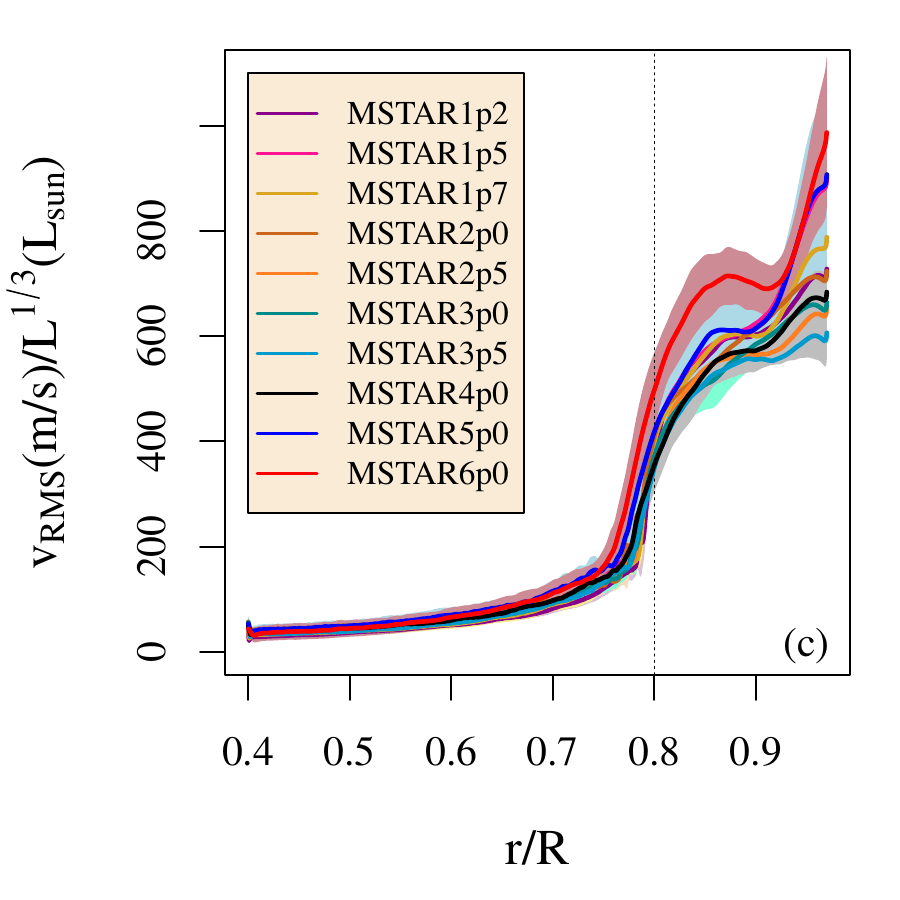}\includegraphics[width=.5\columnwidth]{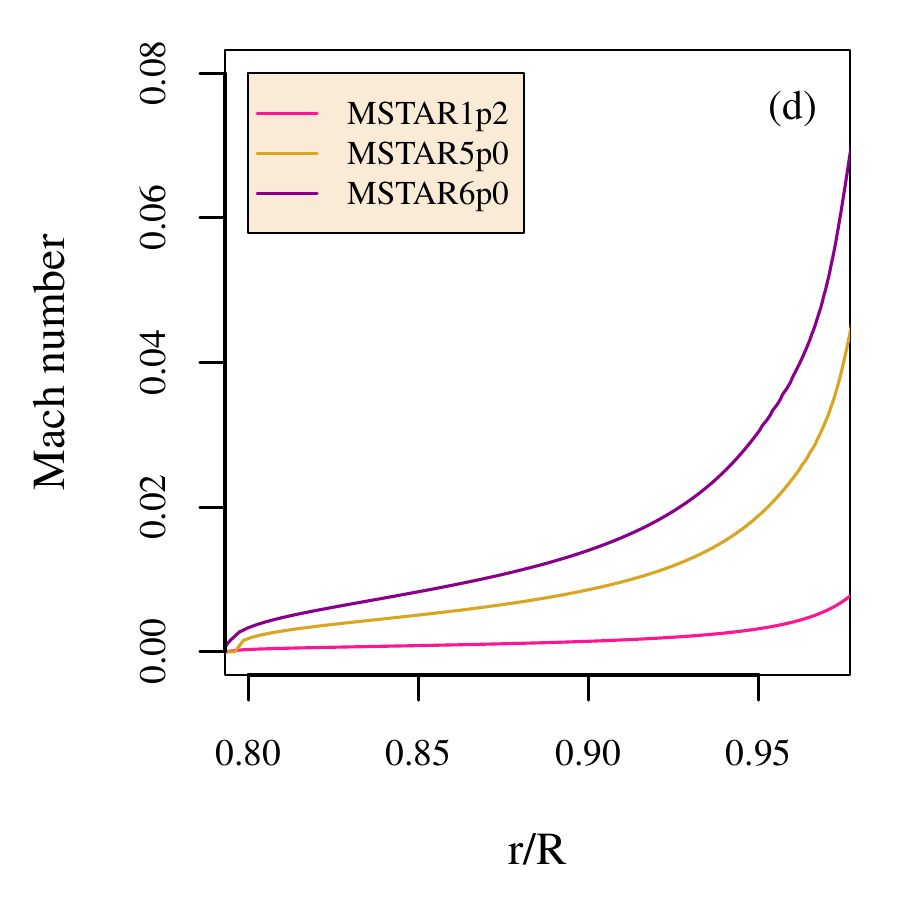}
\caption{ (a) Root-mean-square velocity at the convective boundary $v_{\mathsf{CB}}$ normalized by the cube-root of the luminosity at the surface vs mass.   A thin horizontal line indicates the average value of the normalized $v_{\mathsf{CB}}$ for the eight lowest mass stars examined.  Bars are plotted based on the standard deviation in time for this time-averaged value at the convective boundary.  (b) $v_{\mathsf{CB}}$ vs gravitational acceleration at the photosphere.  A linear regression fit is shown for the log-log plot. (c) the radial profile of the root-mean-square velocity normalized by the cube-root of the luminosity at the surface.  Shaded areas indicate one standard deviation above and below the time-averaged mean of the radial profile. A dotted vertical line indicates the Schwarzschild boundary for convective instability on the radial profile.  (d) Mach numbers predicted by MESA convective velocities. \label{fig:vcbmass}}
\end{figure}

In addition to the convective turnover time, we examine the time-averaged root-mean-square (RMS) velocity profiles of these stars.  The RMS velocity at the convective boundary $v_{\mathsf{CB}}$ normalized by the cube root of luminosity is roughly constant for the mass range and stars that we study (see Figure~\ref{fig:vcbmass}(a)).   The error bars on this figure come from the standard deviation in time, and all of the error bars cross the average line, except for the 6 solar mass star.  The 6 solar mass star is a clear outlier in this data set, with $v_{\mathsf{CB}}$ roughly 30\% larger than the average of the other stars.   The RMS velocity also decays with increasing gravitational acceleration at the photosphere (see Figure~\ref{fig:vcbmass}(b)).  This decay follows an approximate power-law
\begin{eqnarray}
v_{\mathsf{CB}}\sim g_{\mathsf{eff}}^{-0.4854}~.
\end{eqnarray}
The radial profiles of the time-averaged RMS velocity neatly collapse when normalized by the luminosity (see Figure~\ref{fig:vcbmass}(c)), except that the 6 solar mass star's normalized velocity remains higher.  
We examine the profiles of Mach number predicted by the MESA stellar structure, which we define as the convective velocity divided by the sound speed  (see Figure~\ref{fig:vcbmass}(c)).  The 6 solar mass has a considerably higher predicted Mach number than the other simulations, approaching $0.08$ at the outer boundary of the simulation.    The velocities in our hydrodynamic simulations are between 3 and 4 times larger than the convective velocity produced by MLT in MESA.  Taking this into account, the Mach number in our simulations approaches roughly 0.3 near the outer boundary of simulation MSTAR6p0.  This Mach number is challenging, but possible to examine in our fully compressible simulations.
Our brief study of the stellar structures, along with these statistics of convection, demonstrates that the 6 solar mass star has some clear differences from the rest of this series.

\subsection{The Extreme Value Theory analysis of convective overshooting}

We have proposed an improved model for determining the depth of an overshooting or penetration layer, based on a statistical analysis of the depth reached by all convective plumes that penetrate below the large convection zone in a young sun in 2D \citep{pratt2017extreme} and in 3D \citep{pratt2020comparison}.  In the young sun, the convective boundary is in the deep interior where the P\'eclet number is large $\mathsf{Pe} \gg 1$.  Our current series of simulations examines a shallow convection zone where the P\'eclet number may be lower.  Because we do not examine simulation times long enough for the temperature gradients to evolve, and seek only to measure the distance convective plumes reach below the convective region, the distinction between overshooting and penetration is not relevant to this work.  We therefore refer to this process as \emph{overshooting} rather than penetration.

\begin{figure}
  \begin{minipage}[c]{0.7\textwidth}
  \resizebox{4.in}{!}{\includegraphics{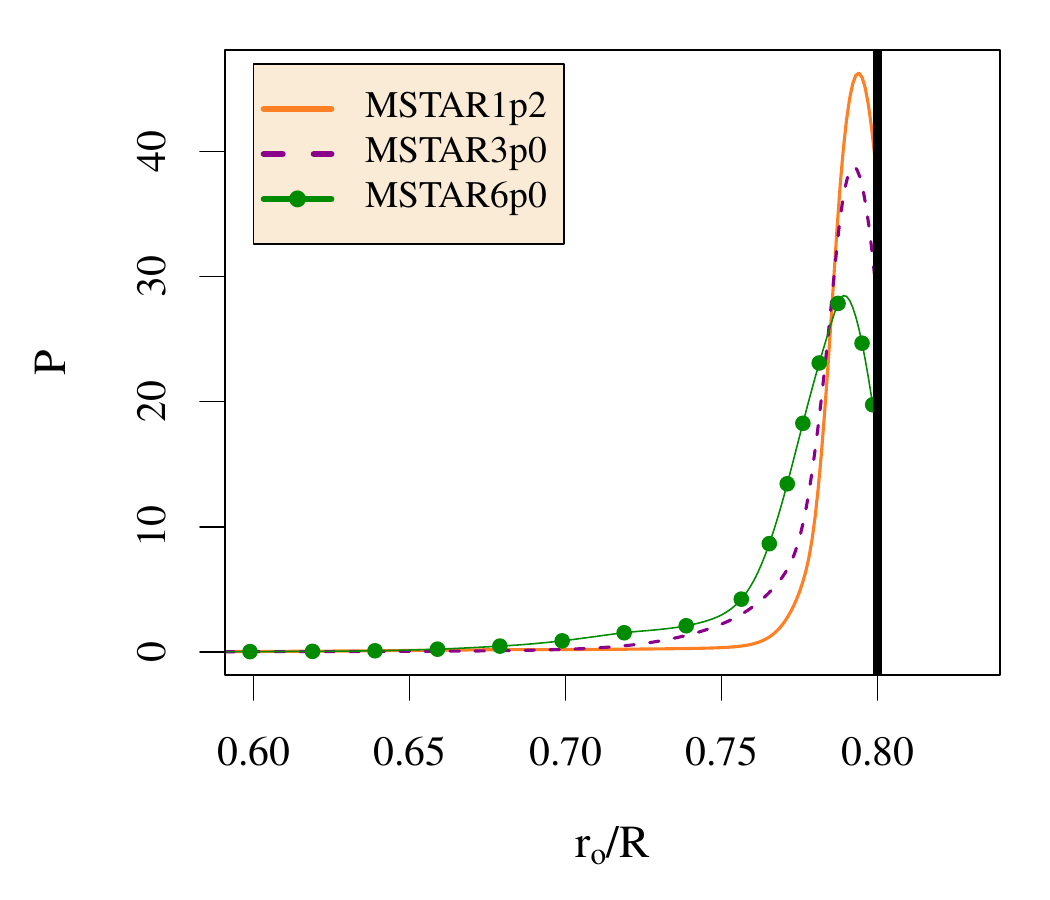}}
  \end{minipage}
  \hspace{5mm}\begin{minipage}[c]{0.25\textwidth}
\caption{Probability density functions of overshooting depth $r_{\mathsf{o}}$ determined by the first zero of the vertical kinetic energy flux, for
the simulations MSTAR1p2, MSTAR3p0, and MSTAR6p0.  The vertical line indicates the bottom of the convection zone determined by the Schwarzschild criterion.
\label{figkpdf}}
  \end{minipage}
\end{figure}

\begin{figure*}
\begin{center}
\resizebox{5.5in}{!}{\includegraphics{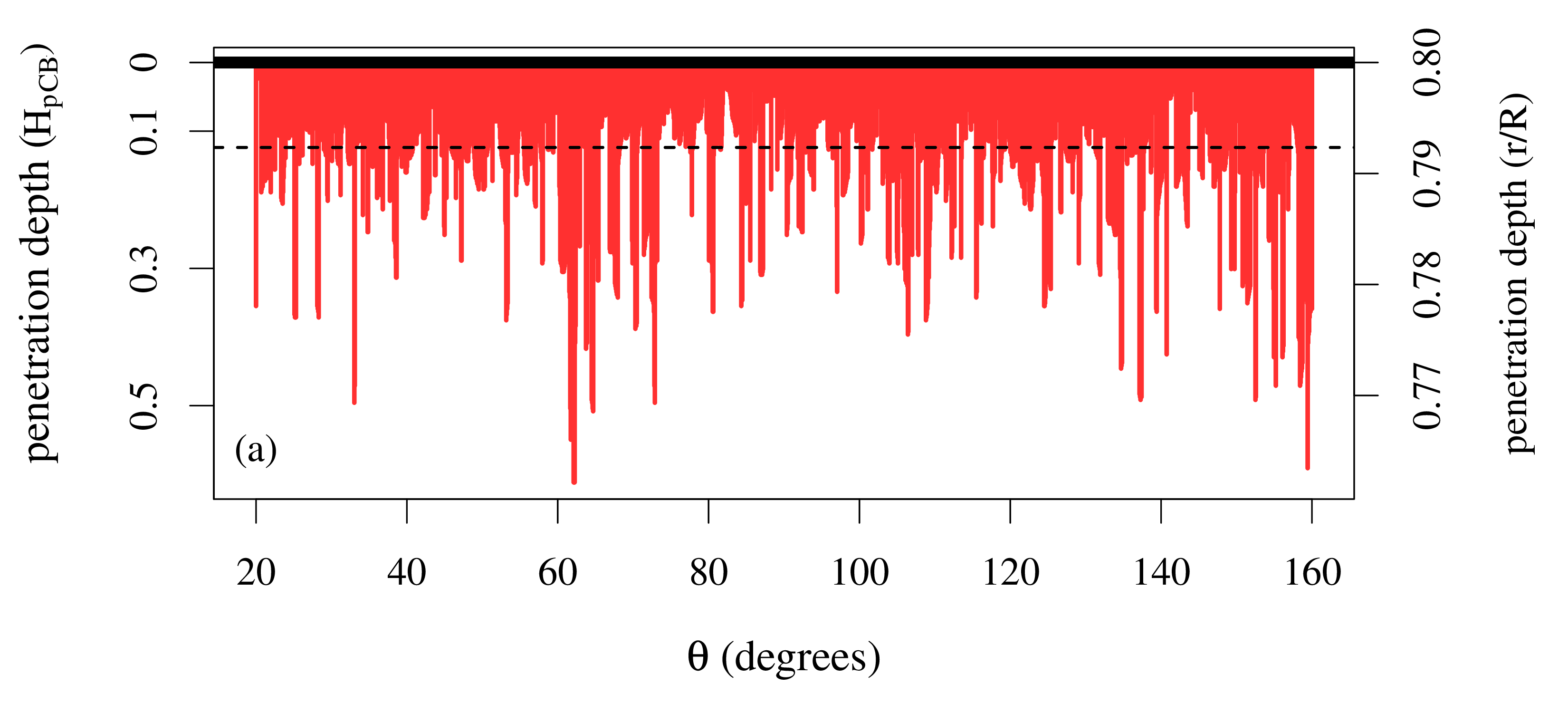}}
\resizebox{5.5in}{!}{\includegraphics{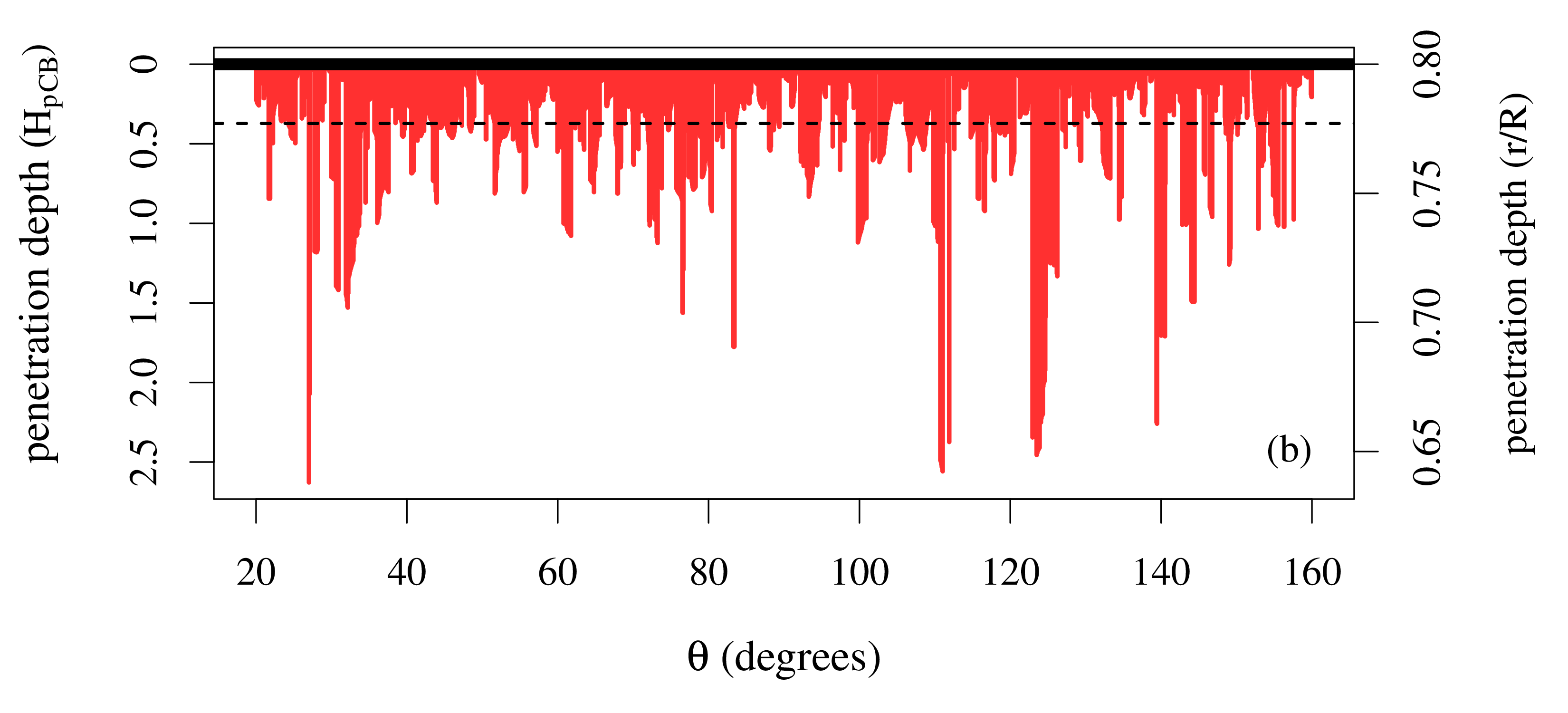}}
\caption{Angular structure of the penetration layer at an arbitrary time in simulation (a) MSTAR1p2 and (b) MSTAR6p0.  The penetration depth in this illustration is determined by the first zero of the vertical kinetic energy flux.  The boundary between the convection zone and the stable radiative zone, calculated from the Schwarzschild criterion, is indicated by a solid black line.   The vertical axis is in units of the pressure scale height $H_{p\mathsf{CB}}$ at this boundary.  A dashed black line indicates the average penetration depth at this time.
\label{figred}}
\end{center}
\end{figure*}

Our model for the depth of the overshooting layer rests on the observation that the statistics of overshooting lengths, calculated for each angular grid cell at each time step in our simulation data, produce a strongly non-Gaussian probability distribution, in which the wings of the distribution are of primary physical importance.   This distribution is essentially non-Gaussian by construction; an overshooting distance can never be negative.  The convective boundary itself cuts off the lower wing of the distribution, so that it can never be symmetrical about its mean value.  The use of an average quantity also removes critical information about the intermittency of convective penetration.   For stars that have more vigorously overshooting plumes, and for higher resolution simulations of stars, the distribution of overshooting plumes becomes skewed in a prominent way.  This is clear in the present simulations (see Figure~\ref{figkpdf}) where the higher mass simulation MSTAR6p0 has a much higher wing than lower mass simulations MSTAR3p0, or MSTAR1p2.  In the figure, the peaks of the distributions are at a similar position below the convective boundary, and the shoulders and wings of the distribution define the differences in the overshooting layers in these stars.  A simple average in both angle and time has been frequently used in early works on this topic, and poorly characterizes the differences between these distributions. Instantaneous examples of the structure of the overshooting layer for the present simulations are illustrated in Figure~\ref{figred}. 

In \citet{pratt2017extreme} we found that when different theoretical measures (vertical kinetic energy flux or vertical heat flux) are used to calculate the point where descending plumes cease, the full probability distributions of penetration depth are similar in form, but the averages can be different.  In that work, we proposed that instead of an average, the \emph{maximum} depth of plume penetration calculated at a single time should be used to define the width of the penetration layer.  The maximum depth of penetration suggests the application of extreme value theory \citep{castillo2005extreme,charras2013extreme,gomes2015extreme} to determine a form for the diffusion coefficient, enhanced by large-scale convective mixing, to model penetration.  The form of this enhanced diffusion coefficient is:
\begin{eqnarray}\label{newEVTdiffusion}
D(r) = D_0 \mathsf{Pe_{B}}^{1/2} \left( 1- \exp{\left( - \exp{\left(-   \frac{(r_{\mathsf{B}}-r)/R - \mu}{ \lambda} \right)} \right)}  \right)~.
\end{eqnarray}   
Here, $\mathsf{Pe_{B}}$ is a characteristic P\'eclet number for the bottom of the convection zone, $r_{\mathsf{B}}$ is the radial position of the convective boundary, and the constants $\lambda$ and $\mu$ are the scale parameter and location parameter of the generalized extreme value distribution (GEVD), which best fits the penetration depth statistics.  Our enhanced diffusion coefficient was analyzed and applied for stellar evolution calculations \citep{baraffe2017lithium,jorgensen2018addressing,dietrich2018penetrative,augustson2018penetration}.  We refer the reader to \citet{pratt2017extreme} for a complete examination of these statistics and discussion of the development of this model.   The data that was used to develop this enhanced diffusion coefficient were from simulations of a young sun with a deep convection zone and one solar mass.  Here, we apply the model to examine convective overshooting for stars that have a shallow convection zone, and are more massive than the young sun.  We fit the long time series of steady-state convection data from each of our simulations to this statistical model, and associate the location parameter $\mu$ with the overshooting depth $\ell_{\mathsf{ov}}$.
\begin{figure*}
\includegraphics[width=.45\columnwidth]{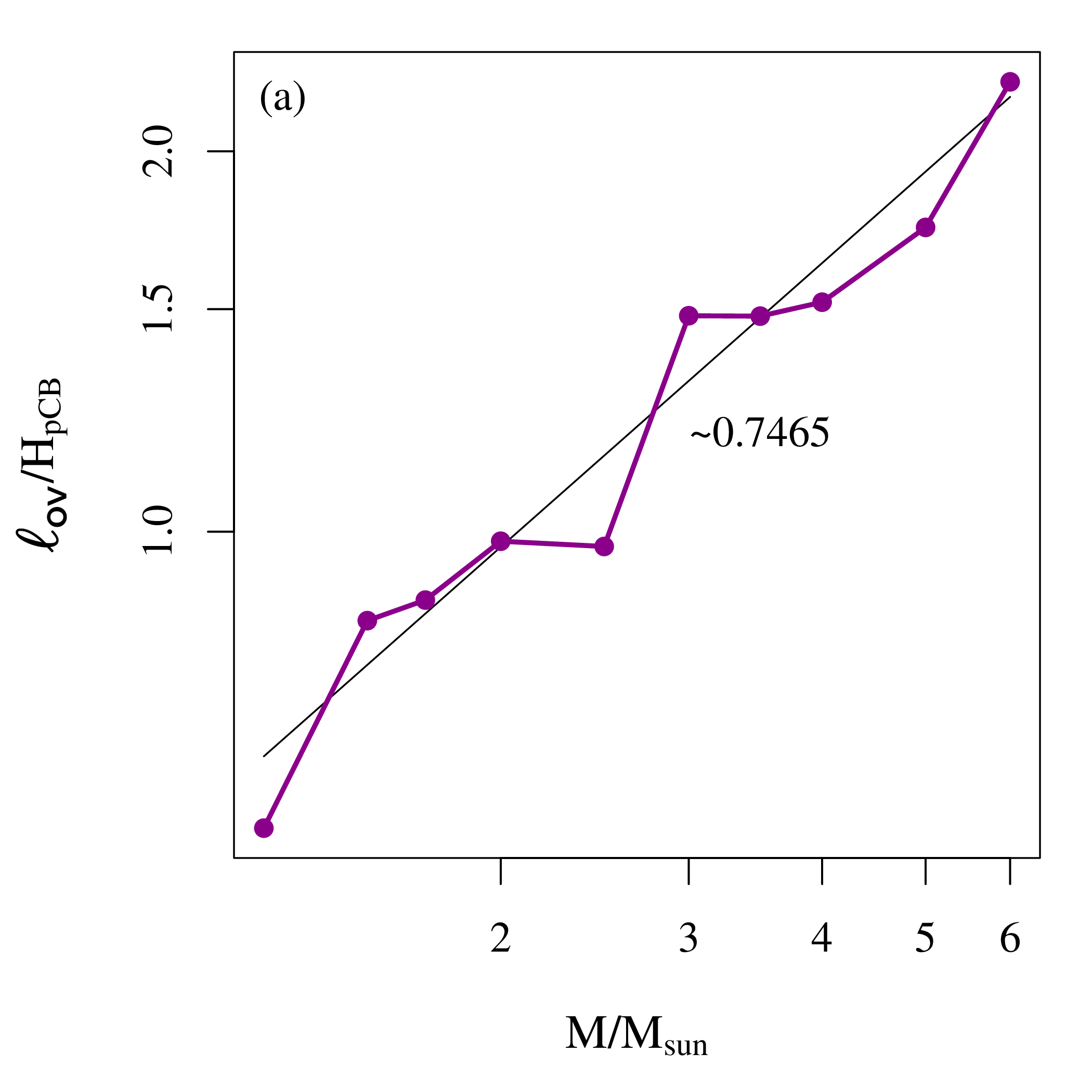}\includegraphics[width=.45\columnwidth]{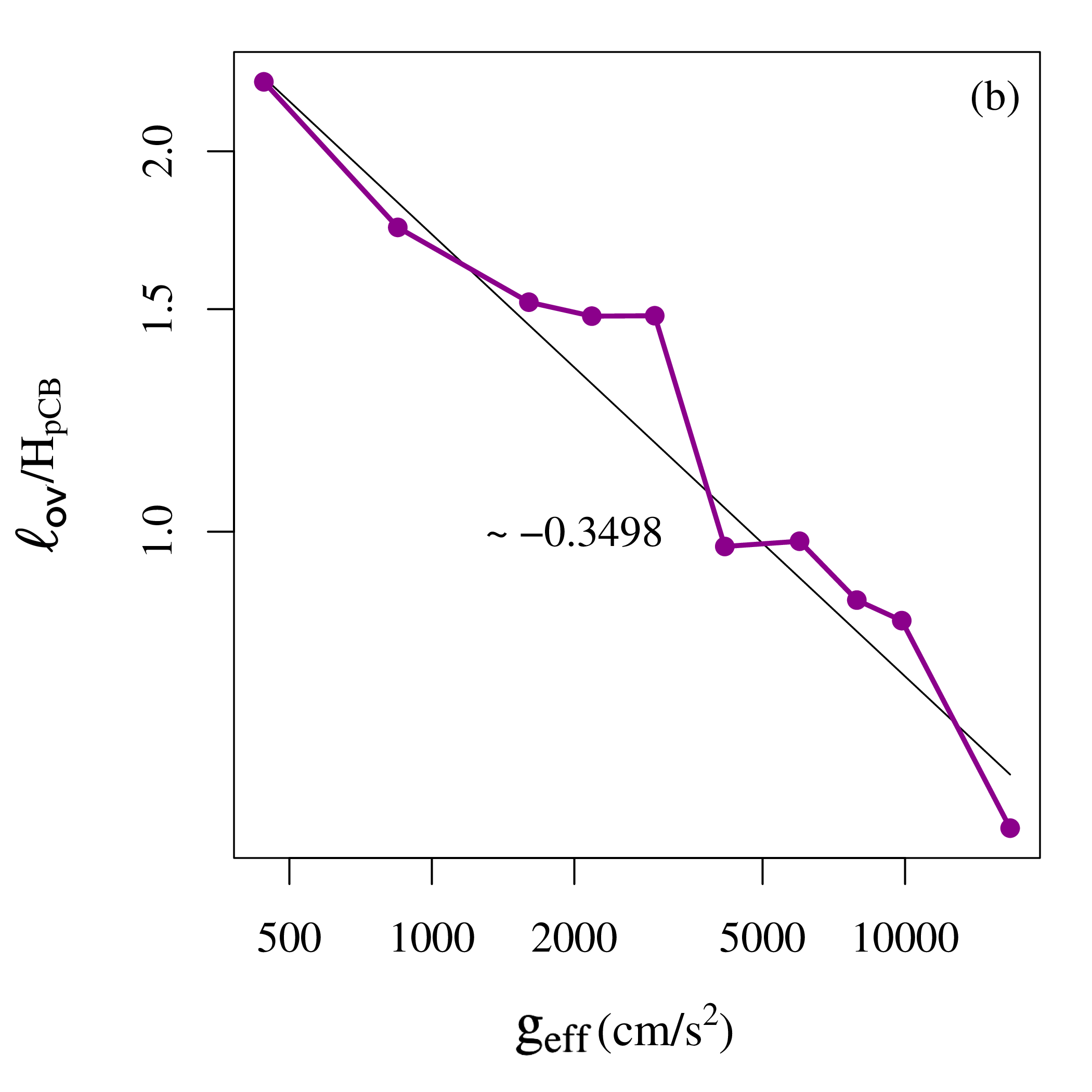}
\caption{Overshooting depth $\ell_{\mathsf{ov}}$ for the ten simulations in Table~\ref{tab:musicparams}.  Linear regression fits are shown for each plot.  \label{fig:lov}}
\end{figure*}

Many stellar evolution calculations assume that the overshooting depth $\ell_{\mathsf{ov}}$ is simply a fixed percentage of the pressure scale height at the convective boundary $H_{p\mathsf{CB}}$, and the pressure scale height will change with the stellar mass or the gravitational acceleration.  In Figures~\ref{fig:lov} this assumption is decisively violated for the stars examined in this work.  Between a mass of $1.2 M_{\mathsf{sun}}$ and  $6.0 M_{\mathsf{sun}}$ the overshooting depth $\ell_{\mathsf{ov}}$ varies from $0.58 H_{p\mathsf{CB}}$ to $2.27 H_{p\mathsf{CB}}$, nearly a factor of 4 change. 

\begin{figure*}
\includegraphics[width=.45\columnwidth]{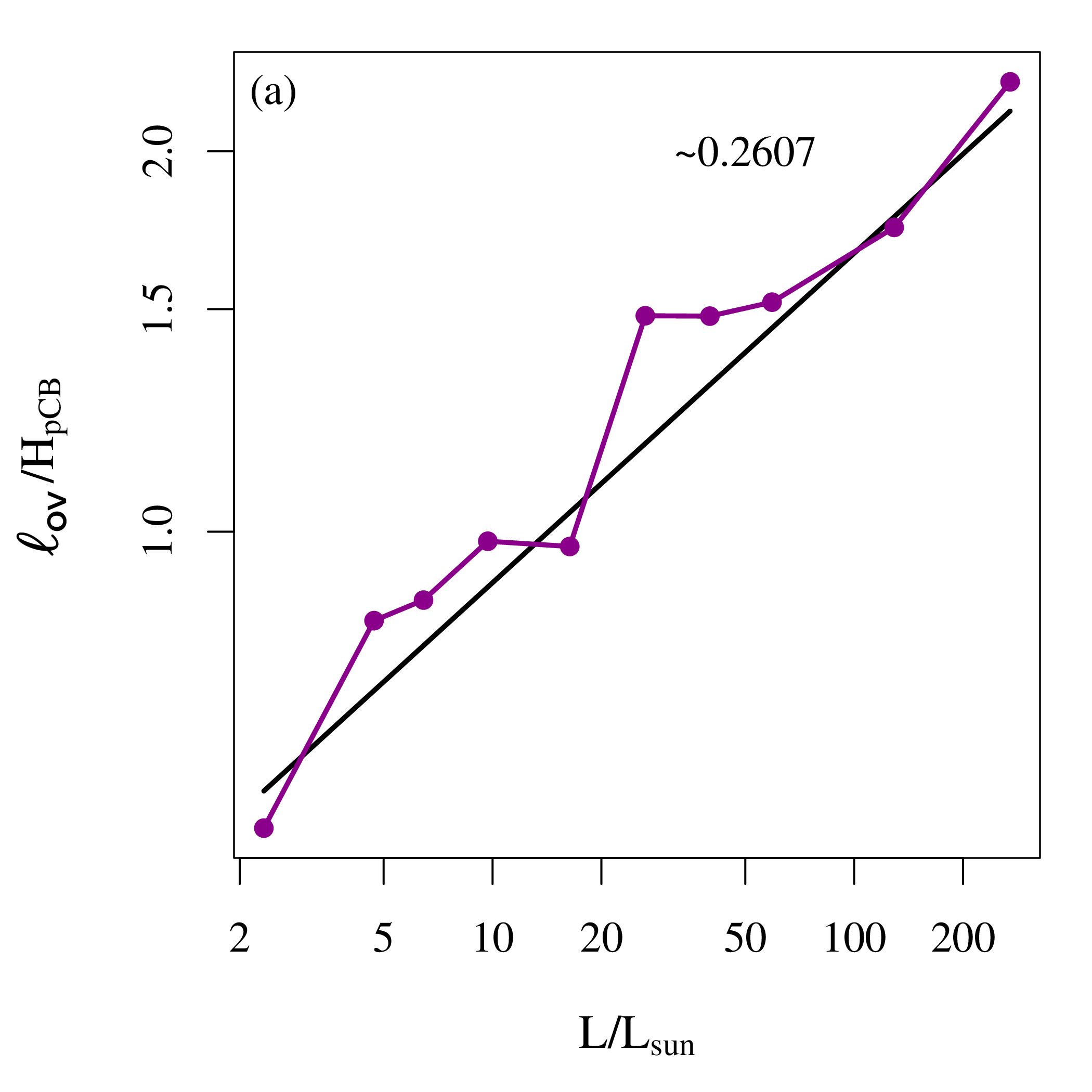}\includegraphics[width=.45\columnwidth]{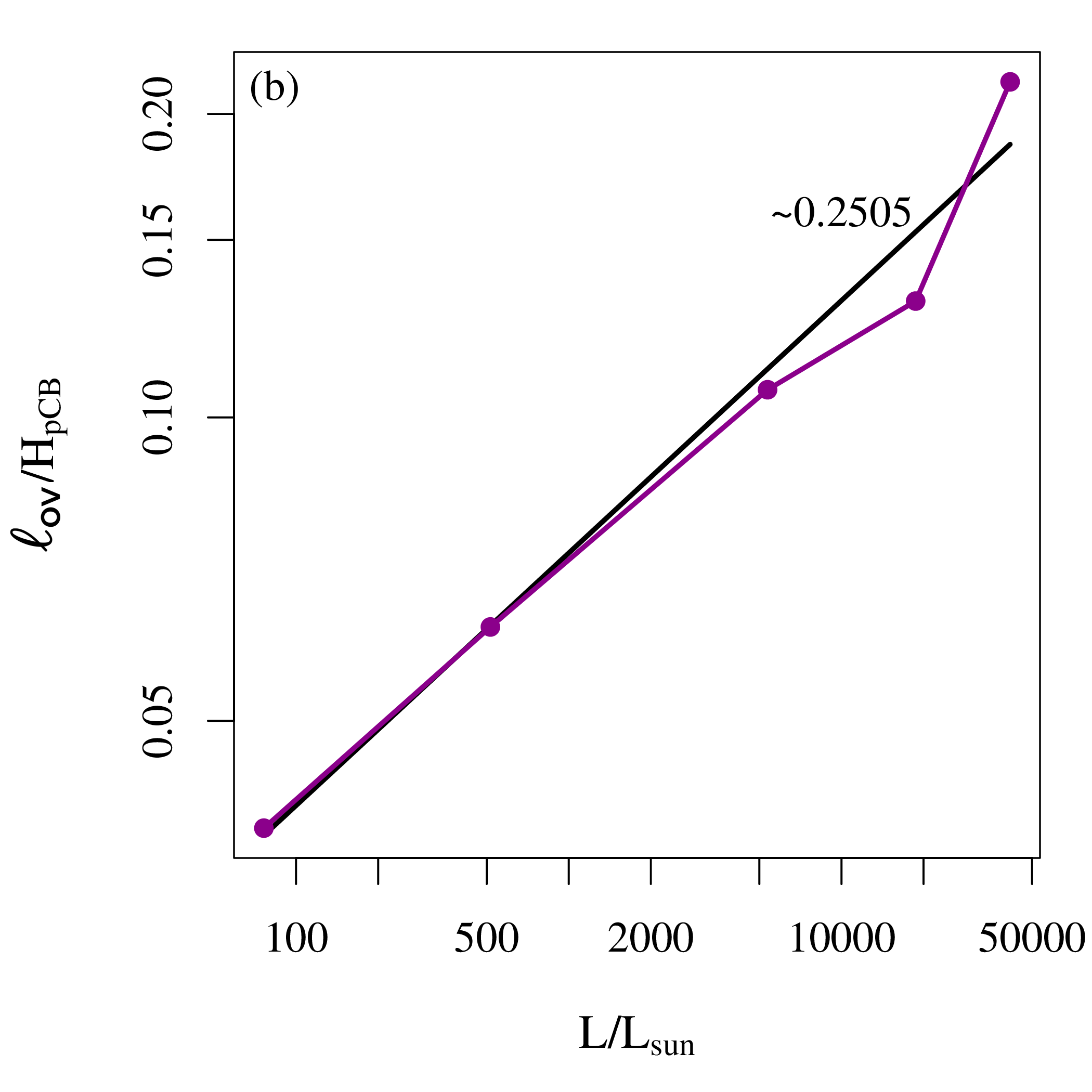}
\caption{Overshooting depth $\ell_{\mathsf{ov}}$ vs luminosity for (a) the ten convective envelope simulations in Table~\ref{tab:musicparams}, and (b) the five core convection simulations studied in  \citet{baraffe2023study}.  Linear regression fits are shown for each plot.  \label{fig:lovlum}}
\end{figure*}

Examining five convective core simulations that span a large luminosity range, \citet{baraffe2023study} noted a scaling of the overshooting depth with the luminosity of
\begin{eqnarray}\label{baraffe23}
\ell_{\mathsf{ov}}/H_{p \mathsf{CB}} \sim (L/L_{\odot})^{1/3} (R_{\mathsf{ext}}/H_{p \mathsf{CB}})^{1/2}~. 
\end{eqnarray}
In this scaling relation, the dependence on the extent of the convective region $R_{\mathsf{ext}}$ echos the theoretical scaling relation that \citet{zahn1991convective} developed for convective cores (see  eq. (4.5) in that work); we will refer to Zahn's scaling relation for convective cores as Z91c and the scaling of \citet{baraffe2023study} as B23.   A second theoretical scaling relation was provided by \citet{zahn1991convective} for convective envelopes (see eq. (3.9) in that work), namely
\begin{eqnarray}\label{zahntotal}
\ell_{\mathsf{ov}} = H_{p\mathsf{CB}} v_{\mathsf{CB}}^{3/2}  (cf)^{1/2}  g_{\mathsf{eff}}^{-1/2} \left[\frac{3}{2} Q K \chi_p \nabla_{\mathsf{ad}} \right]~.
\end{eqnarray}
Here $f$ is called the filling factor, $c$ is the asymmetry parameter, $Q$ is the expansion coefficient at constant pressure, $K$ is the thermal diffusivity, $\chi_p$ is the adiabatic log derivative of radiative conductivity with respect to pressure, and  $\nabla_{\mathsf{ad}}$ is the adiabatic log derivative of temperature with respect to pressure. 
We will refer to \citet{zahn1991convective}'s scaling relation for convective envelopes in eq.~\eqref{zahntotal} as Z91e.
 Unlike convective cores, Z91e does not include any dependence on the radial extent of the convective layer $R_{\mathsf{ext}}$.
 The difficulty that this presents is that luminosity is not explicitly included in Z91e, and other quantities in addition to the velocity can depend on the luminosity.
 Using an approximate scaling of $v_{\mathsf{CB}} \propto L^{1/3}$, and assuming that no other quantity in Z91e is dependent on luminosity, this would indicate
 \begin{eqnarray}\label{lumscaling}
 \ell_{\mathsf{ov}} \propto L^{1/2} ~.
 \end{eqnarray}
For the ten simulations that we study in this work, we observe a scaling exponent that is closer to $1/4$ (see Figure~\ref{fig:lovlum}).   Looking further, the effective gravity scales strongly with the luminosity (see Figure~\ref{fig:loggmass}) with an exponent of about $-0.7$; using this value in Z91e would contribute a factor of $L^{0.35}$.  However, like the mass-luminosity relations, such scalings will likely depend on the evolutionary state of the stars considered.  Including the luminosity dependence of the effective gravity for the current series of pre-main-sequence stars does not account for the scaling of $1/4$ that we observe.

\begin{figure}
  \begin{minipage}[c]{0.6\textwidth}
  \resizebox{4.in}{!}{\includegraphics{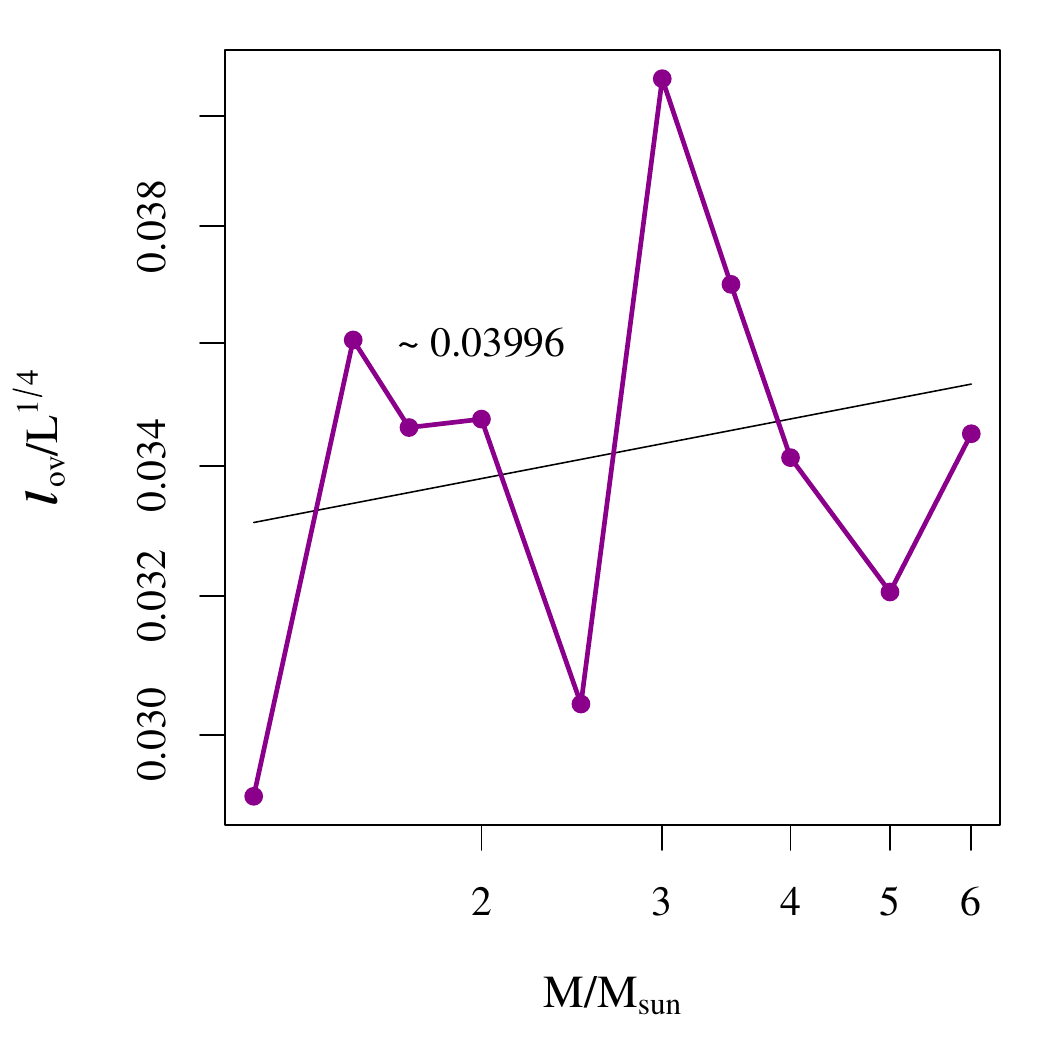}}
  \end{minipage}
  \hspace{5mm}\begin{minipage}[c]{0.3\textwidth}
\caption{Overshooting depth $\ell_{\mathsf{ov}}$, scaled by the luminosity to the $1/4$, vs mass.  A linear regression fit is performed on the log-log plot. \label{fig:lovscaled}}
  \end{minipage}
\end{figure}
We therefore offer the scaling for the rise of the overshooting depth for the present stars:
\begin{eqnarray}\label{suggestedscaling}
\frac{\ell_{\mathsf{ov}} }{H_{p \mathsf{CB}} }\propto \left(\frac{L}{L_{\mathsf{sun}}} \right)^{1/4} ~.
\end{eqnarray}
Here the pressure scale height has been included so that quantities are nondimensionalized.
The slope of the overshooting depth normalized by this luminosity scaling in Figure~\ref{fig:lovscaled} is smaller than the variation of the data, indicating that this luminosity scaling is a good fit to our simulation data; nondimensionalizing these quantities does not modify this.
Eq.~\eqref{suggestedscaling} agrees closely with the data-driven convective core scaling of eq.~\eqref{baraffe23}.  Not accounting explicitly for the dependence on $R_{\mathsf{ext}}$ in eq.~\eqref{baraffe23}
results in a reduction of the luminosity scaling exponent to ${1/4}$, as we find for the stars with convective envelopes (see Figure~\ref{fig:lovlum}). 

\subsection{The predictive power of the filling factor for convection}

The Z91e scaling suggests that a filling factor for convection plays a role in determining the overshooting depth.   As noted by \citet{dethero2024shape}, the theoretical basis for Z91e neglects some of the more complex fluid dynamics that occurs near the convective boundary, e.g. boundary layers, or interactions between waves and convective flows.   As a result the form for a filling factor that Zahn suggests is not viable for the analysis of data from global simulations of stars.

\begin{figure}
  \resizebox{3.3in}{!}{\includegraphics{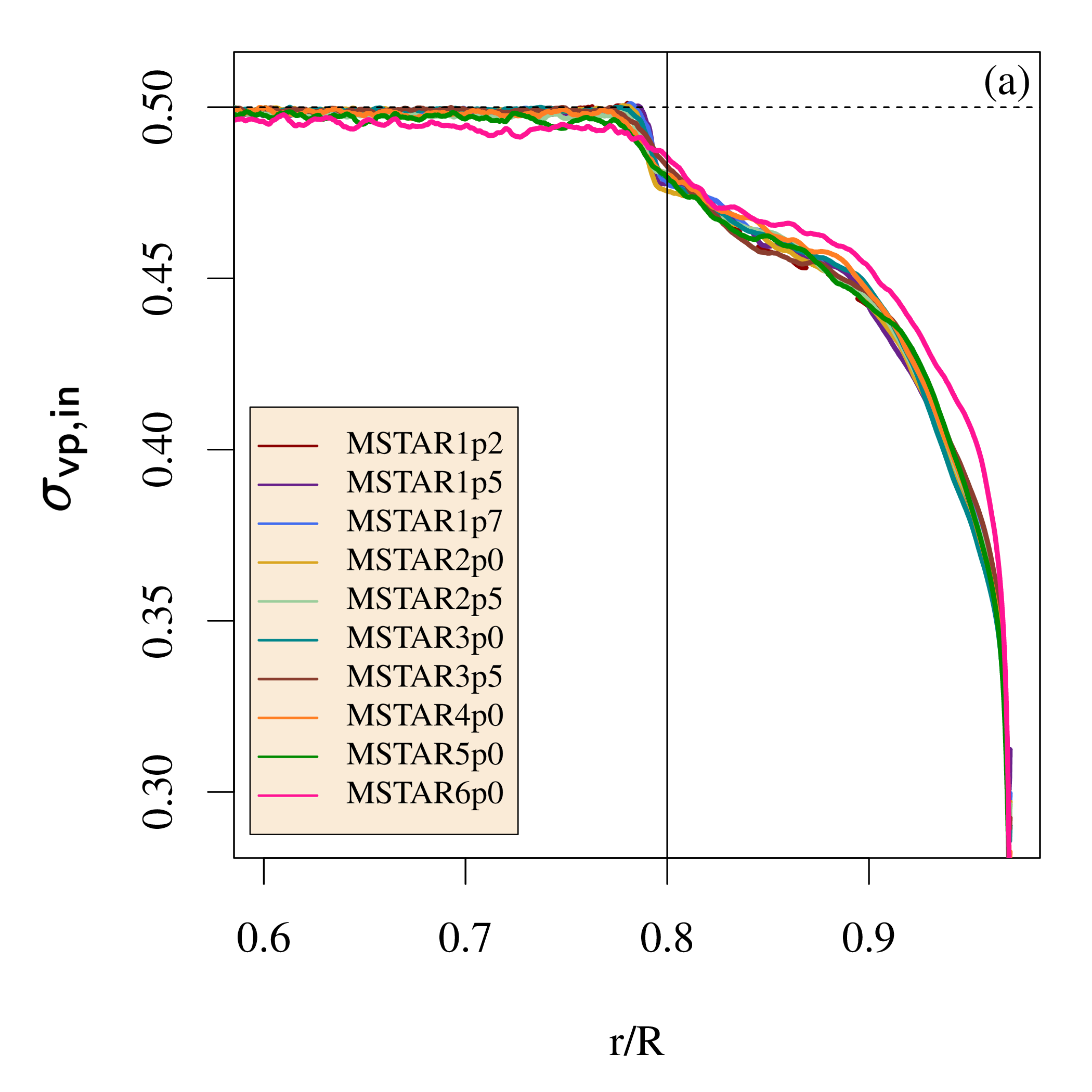}}\resizebox{3.3in}{!}{\includegraphics{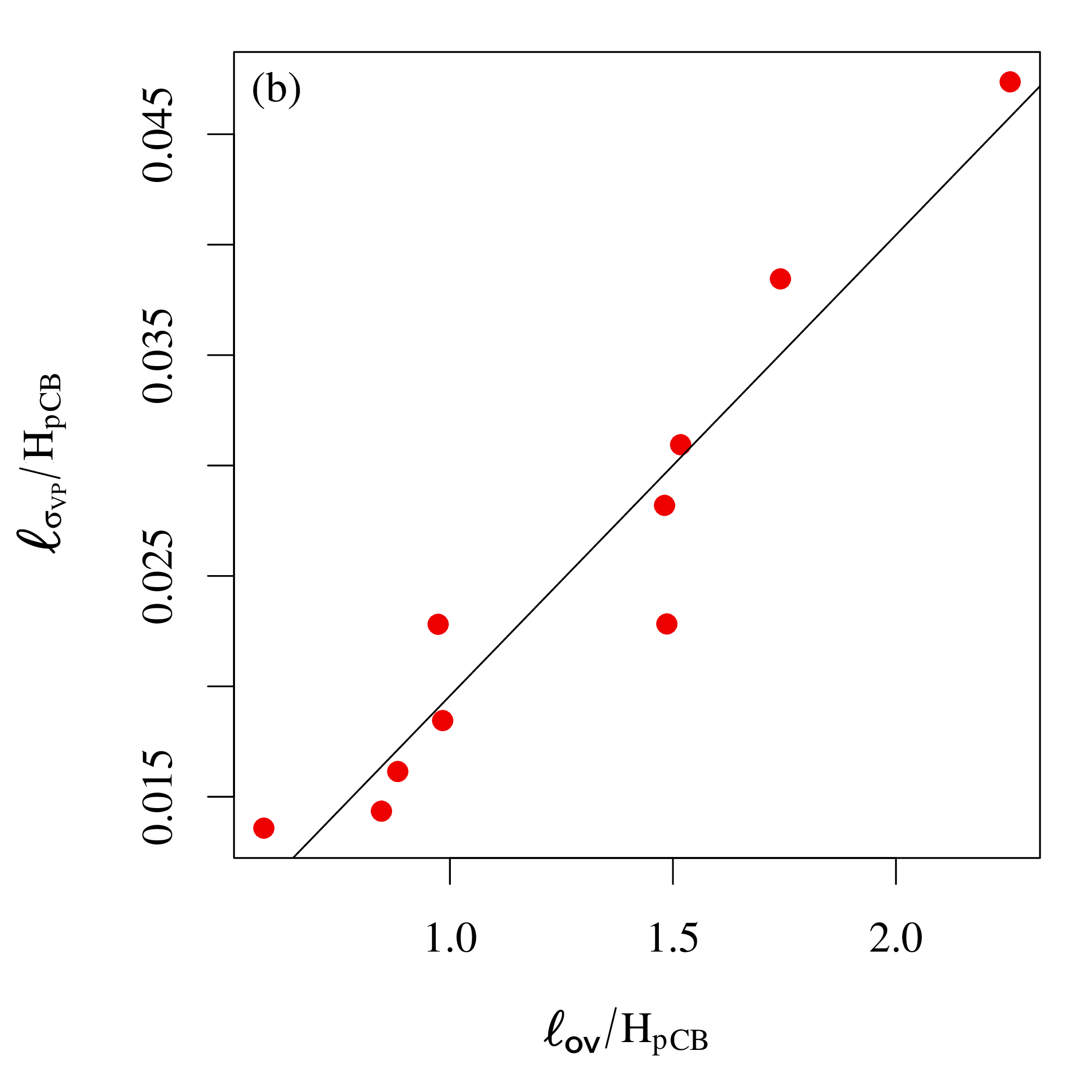}}
\caption{(a) Radial profile of the time-averaged volume-percentage filling factor.  The vertical black line indicates the convective boundary calculated by the Schwarzschild criterion. (b)  The asymmetry depth defined from the volume-percentage filling factor.  Data is shown for the ten simulations in Table~\ref{tab:musicparams}.
\label{fig:sigma}
}
\end{figure}

A simple measure of a filling factor can be defined as the fraction of volume at a given radius occupied by inflows, where inflows are defined by their radial velocity:
\begin{eqnarray}\label{vpfffirst}
\sigma_{\mathsf{vp,in}} &=& \frac{ V^{\mathsf{in}}(r,\theta,t)}
{ V^{\mathsf{in}}(r,\theta,t) + V^{\mathsf{out}}(r,\theta,t) }~.
\end{eqnarray}
For all of the stars studied in this work, the radial profile of the volume-percentage filling factor is very similar (see Figure~\ref{fig:sigma}(a)). At the surface of every simulation  $\sigma_{\mathsf{vp,in}} \approx 0.3$, indicating that inflows are narrow compared to outflows; this is in general agreement with the observed asymmetry at the solar surface.   At the convective boundary, the filling factor for every simulation is $\sigma_{\mathsf{vp,in}} \approx 0.47$, such that inflows and outflows are nearly symmetrical.   For our current series of stars, $\sigma_{\mathsf{vp,in}}$ is highly similar regardless of mass or luminosity.  As we found for the velocity statistics, small differences are evident between the 6 solar mass simulation and the rest.
Although the volume-percentage filling factor is simple to calculate, it does not provide further understanding of the overshooting depth.  Its value lies in providing a comparative measure of the anisotropy of the convection.  

We therefore try to use the slight anisotropy of convective flows to extract information about overshooting plumes.  In the radiative zone, the time-averaged volume-percentage filling factor $\sigma_{\mathsf{vp,in}} \approx 0.5$.  This point is reached shortly below the convective boundary.  We define a new length scale based on the distance from the convective boundary to the point where $\sigma_{\mathsf{vp,in}} \approx 0.5$.   We call this new length scale the asymmetry depth $\ell_{\sigma_{\mathsf{vp}}}$. A linear scaling provides a good fit between the overshooting depth calculated by our EVT analysis and the asymmetry depth (see Figure~\ref{fig:sigma}(b)).     The asymmetry depth thus agrees with the scaling for overshooting depth, but provides no new insight into how overshooting should scale with other fundamental parameters such as mass and luminosity.

\begin{figure}
  \resizebox{3.3in}{!}{\includegraphics{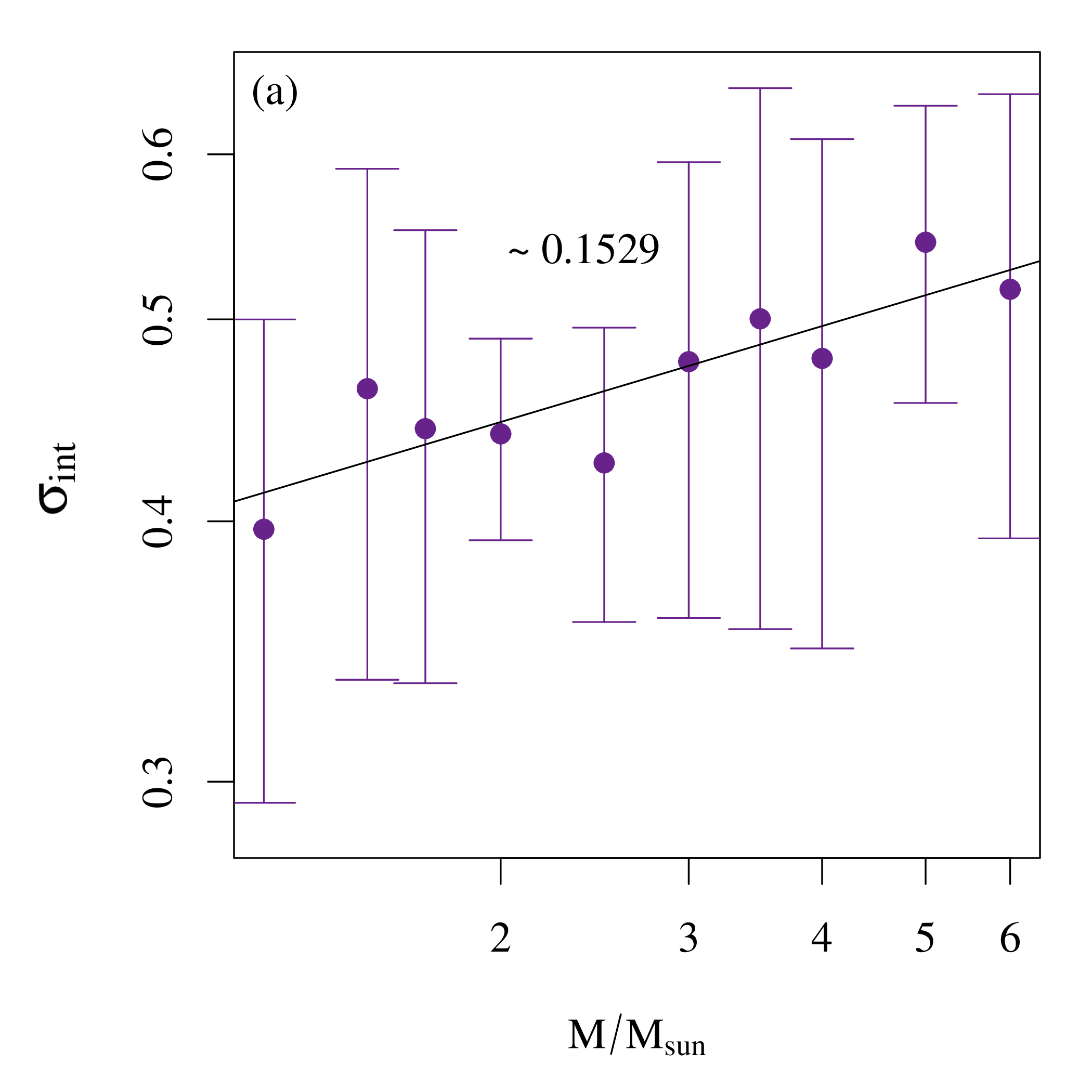}}\resizebox{3.3in}{!}{\includegraphics{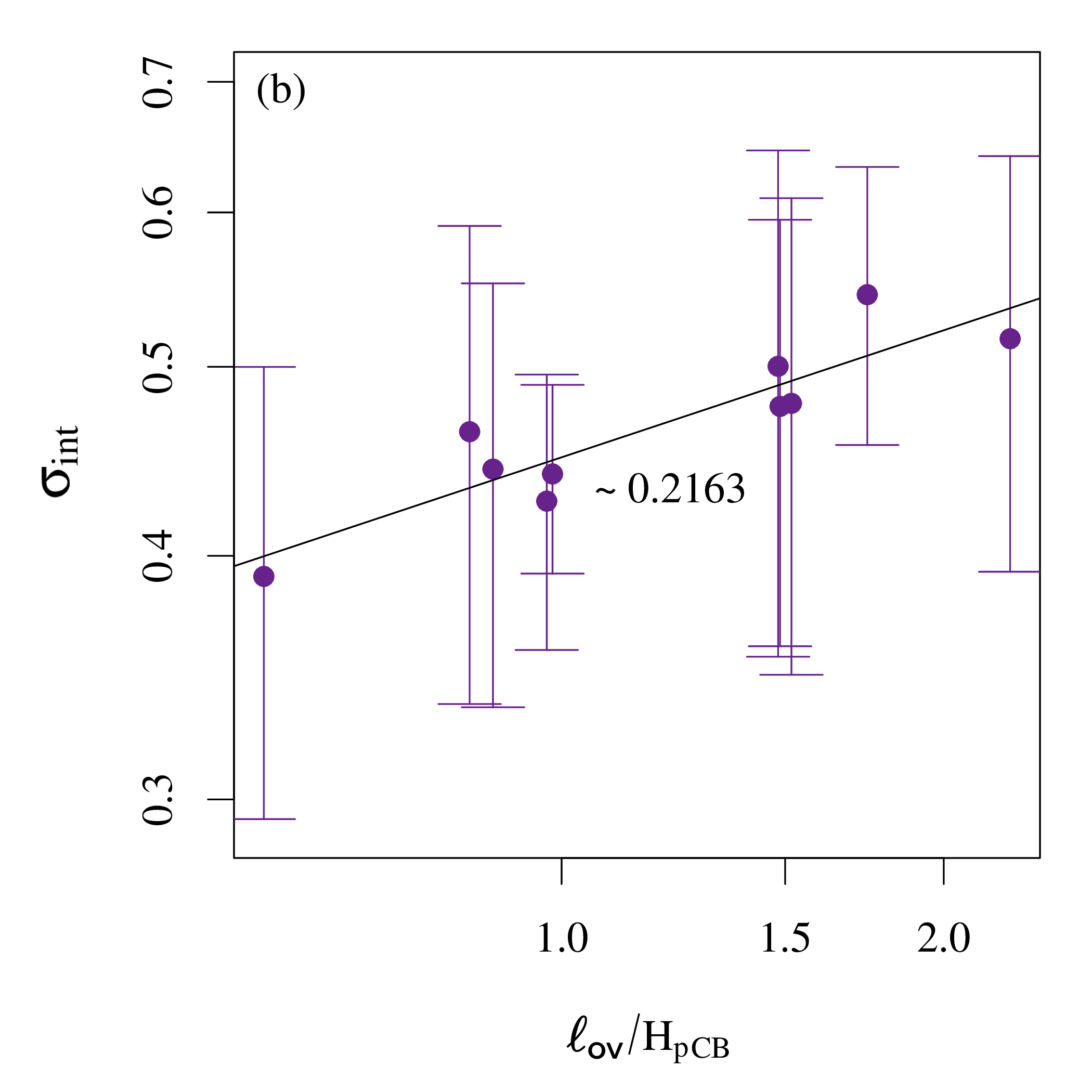}}
\caption{The plume interaction parameter for the ten simulations in Table~\ref{tab:musicparams}: (a) trend vs mass, (b) trend vs overshooting depth.  Linear regression fits are shown on both log-log plots. Error bars represent one standard deviation in time from the time-averaged point. \label{fig:lovsigmaint}
}
\end{figure}

\begin{figure}
  \begin{minipage}[c]{0.6\textwidth}
  \resizebox{3.5in}{!}{\includegraphics{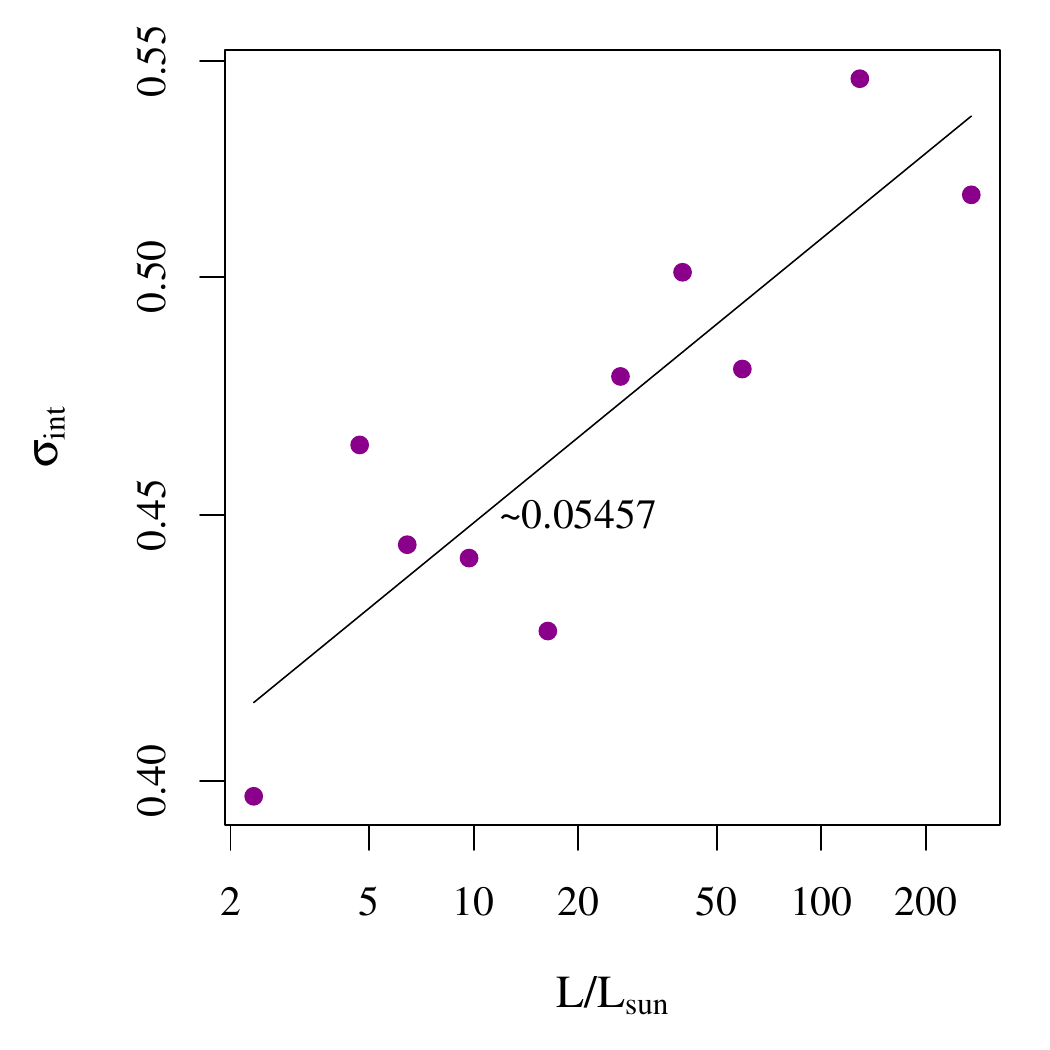}}
  \end{minipage}
  \hspace{5mm}\begin{minipage}[c]{0.3\textwidth}
\caption{The plume interaction parameter vs luminosity.  A linear regression fit is shown on the log-log plot.  Errors on these points are identical to those shown in Figure~\ref{fig:lovsigmaint}. \label{fig:lovsigmaintlum}}
  \end{minipage}
\end{figure}

A second possible filling factor-like quantity is the plume-interaction parameter $\sigma_{\mathsf{int}}$ proposed by \citet{dethero2024shape}.  This is a two-point statistic, defined by the ratio
\begin{eqnarray}\label{defplumeint}
\sigma_{\mathsf{int}} = W_{\mathsf{OL}}/W_{\mathsf{CZ}}~,
\end{eqnarray}
where $W$ is the average width of convective inflows at a given radius,  $\mathsf{OL}$ is the radial point in the overshooting layer where this width profile has a minimum, and $\mathsf{CZ}$ is the radial point in the convection zone where the average width is maximum.  This parameter encapsulates the changing shape of convective inflows around the convective boundary in a simple, zeroeth-order way; it thus fulfills the basic concept of a filling factor, with the twist that it is dependent on data at two radial points.  We speculate that a two-point statistic is better able to capture the non-local aspect of compressible stellar convection than a one-point statistic.  The plume interaction parameter $\sigma_{\mathsf{int}}$ increases with stellar mass and with overshooting depth (see Figure~\ref{fig:lovsigmaint}).  This result demonstrates that for stars in a carefully controlled sequence,  $\sigma_{\mathsf{int}}$ is predictive of the overshooting depth.
The plume interaction parameter $\sigma_{\mathsf{int}}$ also increases shallowly with luminosity (see Figure~\ref{fig:lovsigmaintlum}).  This indicates that it could contribute to the scaling relation in eq.~\eqref{suggestedscaling}.

\section{Summary and Conclusions \label{secconc}}

We have selected MESA models for ten pre-main sequence stars ranging in mass from $1.2 M_{\mathsf{sun}}$ to $6 M_{\mathsf{sun}}$; this mass range was selected because the structures of these stars are very similar, apart from quantities that are directly linked to their mass (e.g. luminosity, radius, gravity).  The particular stellar structures were then selected to have identical aspect ratios, with convective envelopes that cover the outer 20\% of the stellar radius, so that $R_{\mathsf{CB}}=0.8R$.  We perform 2D hydrodynamic simulations with the \music code to quantify the dependence of overshooting depth on mass and luminosity for these stars.

For the overshooting depth, we find the scaling exponent on the luminosity to be $1/4$; there is no clear theoretical reasoning behind the choice of this exponent.  A theoretical scaling developed by \citet{zahn1991convective} (Z91e) does not shed new light on the dependence of overshooting depth on mass or luminosity measured in our simulations.  This is largely because the quantities in Z91e are very similar for the simulations that we consider here.  The data-driven scaling law of \citet{baraffe2023study}, developed for convective cores, has similarities with the scaling we observe.  This scaling law is dependent on both luminosity and the radial extent of the convective layer $R_{\mathsf{ext}}$.   The scaling of our simulations with luminosity agrees with the scaling law of \citet{baraffe2023study} if one neglects the explicit dependence on $R_{\mathsf{ext}}$. 
It is not clear why a scaling for convective cores should also apply to convective envelopes, or whether the dependence on the extent of the convective layer should be included for convective envelopes.
Further work is needed to establish and validate this scaling as well as its theoretical basis.  That work should ideally also include stars that have different structures; the highly similar structures of the stars we studied in this work were a strength, but also prevented us from understanding further dependences of the overshooting depth.  The next steps in the larger effort to understand how stellar mass affects convective overshooting are underway.


Because  \citet{zahn1991convective} developed the idea of a filling factor in the context of scaling laws for the depth of convective overshooting, we have also evaluated filling factors for the simulations studied here.  We have examined a volume-percentage filling factor and the plume interaction parameter proposed by \citet{dethero2024shape}.  The radial profile of the volume-percentage filling factor is highly similar throughout the convection zone for every simulation in this work.  This makes it useless for scaling laws, but does confirm that the stellar structures are producing convection that is similar, with some smaller differences beginning to appear at 6 solar masses.  The plume interaction parameter increases shallowly with the overshooting depth, and also with the luminosity.  This correlation is an indication of its predictive power, and makes it a possible candidate for future scaling laws.

\section*{Acknowledgements}
The research leading to these results is partly supported by the ERC grants 787361-COBOM and by the STFC grant ST/Y002164/1.
This work used the DiRAC Complexity system, operated by the University of Leicester IT Services, which forms part of the STFC DiRAC HPC Facility (www.dirac.ac.uk). This equipment is funded by BIS National E-Infrastructure capital grant ST/K000373/1 and STFC DiRAC Operations grant ST/K0003259/1. DiRAC is part of the National E-Infrastructure.  This work also used the University of Exeter local supercomputer ISCA.  Part of this work was performed under the auspices of the U.S. Department of Energy by Lawrence Livermore National Laboratory under Contract DE-AC52-07NA27344.  LLNL-TR-2010799.

\bibliographystyle{unsrtnat}
\bibliography{masscomp}

\appendix 

\appendix

\section{MESA Inlist}

\begin{table*}
\rowcolors{2}{gray!10}{white}
\centering
\caption{Additional parameters calculated at the convective boundary.  The pressure scale height and overshooting depth are calculated from MUSIC simulations.  The remaining parameters come from the MESA stellar structures.
}
\label{tab:moremesaparams}
\begin{tabular}{l|ccccccccc} 
\hline
$M/M_{\mathsf{sun}}$ & {$H_{p\mathsf{CB}}/R$}&   {$\ell_{\mathsf{ov}}/R$}  & {$M_{\mathsf{enc,CB}}/M_{\mathsf{sun}}$} & {$L_{\mathsf{CB}}/L_{\mathsf{sun}}$} &    {$\kappa_{\mathsf{CB}}$}  &    {$T_{\mathsf{CB}}$}  &    {$p_{\mathsf{CB}}(10^{12})$} &    {$g_{\mathsf{CB}}$ (cm/s$^2$)}  \\
\hline
1.2 &  0.062  &    0.03603 &  1.195432  & 2.333438 &   21.85341    &  1207526 &  2.455552 & 26037.3 \\ 
1.5 & 0.062   &   0.05308 & 1.494036  &   19.55151     &   4.715883    & 1019796  &  0.8876333 & 15288.92 \\ 
1.7 & 0.062   &   0.05514   & 1.692446 &   18.51364       &  6.466469     & 970375.9  & 0.6389724 & 12263.81 \\ 
2.0 & 0.063   &    0.06132  &  1.990162 &   16.37379    &  9.744516     & 915379.5  & 0.4057445 & 9288.451 \\ 
2.5 & 0.063   &    0.06120   &  2.48322  &   15.63618    &   16.45505    & 857018  & 0.2675707  & 6453.115 \\ 
3.0 & 0.062   &    0.09229   & 2.978985 &   14.11732     &  26.63516     & 780748.8  &  0.1407063 &  4573.747 \\ 
3.5 & 0.063   &    0.09296 &  3.47175 &   13.30203    &  40.22608     & 729453.8  &   0.08847722 &  3379.871\\ 
4.0 & 0.062   &    0.09469 & 3.966333   &  12.77545     &  59.74594     &  667714.8 &  0.04996672 & 2486.901 \\ 
5.0 & 0.062   &    0.10806 &  4.964244 &   12.0436    &  129.7389     &  530639.5 & 0.01178033 & 1309.152 \\ 
6.0 & 0.062   &    0.139875 &  5.963429 &   11.40544    &  271.0841     &  417561.8 & 0.002742372 & 685.2771 \\ 
\hline
\end{tabular}
\end{table*}

The simple inlist below was used to produce the stars with a convective envelope in this study.

\begin{verbatim}
&star_job

      show_log_description_at_start = .false.
      create_pre_main_sequence_model = .true.

      save_model_when_terminate = .true.
      save_model_filename = 'final.mod'

      change_initial_net = .true.      
      new_net_name = 'o18_and_ne22.net'

/ ! end of star_job namelist

&controls
      max_num_profile_models = 10000
      use_gold_tolerances = .true.

      initial_mass = [see Table]
      initial_z = 0.02
      
      use_Type2_opacities = .true.
      Zbase = 0.02
      
      mixing_length_alpha = 1.9
      log_L_upper_limit = 4.3

      history_interval = 1
      terminal_interval = 10
      write_header_frequency = 10
      which_atm_option = 'simple_photosphere'
      photo_interval = 500
      profile_interval = 1
     
/ ! end of controls namelist
\end{verbatim}

Here we have used the OPAL Type 2 tables \citep{paxton2013modules}, which
include the OPAL radiative opacities
for carbon and oxygen-rich mixtures.  This choice was
made to allow for the stars to be evolved through the red
giant branch into the helium burning phase, as is shown in Figure~\ref{fig:hrdiagram}.  
There is little difference between the structures produced by
the basic OPAL tables and the Type 2 tables
 for the pre-main-sequence stars studied
here.  We include several additional parameters that result from these structures in Table~\ref{tab:moremesaparams}.

\end{document}